\documentclass[fleqn,usenatbib]{mnras}

\usepackage[T1]{fontenc}
\usepackage[normalem]{ulem}

\usepackage{graphicx}	% Including figure files
\usepackage{amsmath}	% Advanced maths commands
\usepackage{amssymb}	% Extra maths symbols
\usepackage{supertabular,booktabs}
\graphicspath{{./}{Figures/}}
\usepackage{newtxtext,newtxmath}
\DeclareRobustCommand{\VAN}[3]{#2}
\let\VANthebibliography\thebibliography
\def\thebibliography{\DeclareRobustCommand{\VAN}[3]{##3}\VANthebibliography}

\DeclareRobustCommand{\VANDER}[3]{#2}
\let\VANDERthebibliography\thebibliography
\def\thebibliography{\DeclareRobustCommand{\VANDER}[3]{##3}\VANDERthebibliography}

\usepackage{xspace}

\newcommand{\FRI}{FR22\xspace}
\newcommand{\FRII}{FR25\xspace}

\newcommand{\Disco}{\texttt{Disco}\xspace}
\title[Eccentric planet-disc interactions]{Pushing the limits of eccentricity in planet-disc interactions}

\author[Fairbairn \& Dittmann]{
Callum W. Fairbairn$^{1}$\thanks{E-mail: cfairbairn@ias.edu}
and Alexander J. Dittmann$^{1}$\thanks{E-mail: dittmann@ias.edu}\thanks{NASA Einstein Fellow}
\\
$^{1}$School of Natural Sciences, Institute for Advanced Study\\
1 Einstein Drive, Princeton 08540, NJ
}

\date{Accepted XXX. Received YYY; in original form ZZZ}

\pubyear{2025}

\begin{document}
\label{firstpage}
\pagerange{\pageref{firstpage}--\pageref{lastpage}}
\maketitle

% Abstract of the paper

\begin{abstract}
Modelling the gravitational interaction between an eccentric perturber and a differentially shearing gas disc is a longstanding problem with various astrophysical applications, ranging from the evolution of planetary systems to the migration of black holes embedded in AGN discs. Recent work has advanced a global, linear, modal approach for calculating the excited wake and the resulting feedback on the perturber's orbital evolution. In this work we perform a complementary suite of targeted hydrodynamic simulations to test this linear framework across a range of disc temperature and density profiles. In particular, we push from circular orbits to highly eccentric trajectories for which the perturber moves supersonically with respect to the background gas. We find remarkable agreement between our simulations and the linear methodology across a range of diagnostics -- lending support to the predicted wake morphologies, complex radial torque density profiles, and torque reversal phenomena, which occur when the eccentricity exceeds the local aspect ratio. In contrast, comparison with previous fitting functions exposes noticeable discrepancies, cautioning against their indiscriminate use in studies which explore a wide range of perturber eccentricities, in varied disc structures. Our simulations also probe the fundamentally nonlinear effects of shock induced angular momentum deposition and coorbital horseshoe drags, which exhibit clear variations with eccentricity. Finally, this careful comparison between linear theory and numerics provides a detailed benchmark for planet-disc interaction problems and therefore we have provided a repository of our linear calculations for use as a rigorous test of future numerical investigations.
\end{abstract}

% Select between one and six entries from the list of approved keywords.
% Don't make up new ones.

%%%%%%%%%%%%%%%%%%%%%%%%%%%%%%%%%%%%%%%%%%%%%%%%%%

%%%%%%%%%%%%%%%%% BODY OF PAPER %%%%%%%%%%%%%%%%%%

% ================================%
\section{Introduction}
\label{sec:intro}
% ================================%

The interaction between a perturbing body and its gaseous environment is a generic problem applicable to a  wide range of astrophysical systems. For example, the gravitational interaction between a massive perturber and surrounding gas mediates the orbital evolution of protoplanets \citep[e.g.][]{GoldreichTremaine_1979}, the migration and trapping of black holes in active galactic nuclei (AGN) discs \citep[e.g.][]{1993ApJ...409..592A,2011PhRvL.107q1103Y} and the common-envelope inspiral of binary systems \cite[e.g.][]{RopkeDeMarco_2023}. Whilst this last example might be modelled using the dynamical friction formalism, arising from gravitationally focused density wakes in a homogeneous medium \citep[][]{Ostriker_1999}, the perturber-disc interaction is complicated by the differential shear flow inherent to rotationally supported Keplerian systems. In such cases, the interaction is mediated by spiral density waves which exchange energy and angular momentum with the perturber.

Whilst the majority of attention has focused on the regime where the perturber is on a circular orbit, there is evidence to suggest that eccentric orbits might also be common. Indeed, observed spiral arm pitch angles in protoplanetary discs are sometimes difficult to explain using a single, circular planet \citep[e.g.][]{MonnierEtAl_2019, UyamaEtAl_2020}. Furthermore, the presence of multiple spiral pattern speeds in SAO 206462 is suggestive of two planets or an eccentric perturber \citep{XieEtAl_2021}. These planet-disc interactions might also connect with the post-disc-dispersal exoplanet demographics, which exhibit a clear eccentric population \citep[e.g.][]{KaneEtAl_2012,XieEtAl_2016,EylenEtAl_2019,DebrasEtAl_2021}. The origin for these eccentricities might owe to a variety of mechanisms as described in the recent review by \cite{Paardekooper2023}, including planet-planet scatterings \citep[e.g.][]{FordRasio_2008,LegaEtAl_2013}, secular or resonant gravitational interactions with the surrounding disc \citep[e.g.][]{Goldreich2003,TeyssandierOgilvie_2016, Ragusa2018} and thermal torques \citep[e.g.][]{Masset_2017,EklundMasset_2017}.

Much of the physics governing planet-disc interactions also sculpts the orbits of stellar-mass objects orbiting within AGN discs.\footnote{ Moreover, as AGN discs are thought to be quite thin, their scale heights thought to be roughly hundredths of their corresponding radii \citep[e.g.,][]{2002apa..book.....F}, even eccentricities on the order of $e\sim10^{-2}$ may cause order-unity modifications to their orbital evolution \citep[e.g.][]{FairbairnRafikov_2025}.} Disc-perturber interactions drive the migration of objects through the disc \citep[e.g.,][]{BellovaryEtAl_2016,2020MNRAS.493.3732D,2023MNRAS.521.4522D}, aid the formation of disc-embedded binaries \citep[e.g.,][]{2020MNRAS.498.4088M,2020ApJ...898...25T}, and will affect the gravitational wave signals emitted from extreme-mass-ratio inspirals \citep[e.g.,][]{2011PhRvL.107q1103Y,2022MNRAS.517.1339G}. Given a population of embedded objects, collisions may easily excite eccentricities in excess of the accretion disk aspect ratio \citep[e.g.,][]{2017MNRAS.464..946S}, but the relevance of such excitation depends strongly on the rate of eccentricity damping mediated by perturber-disc interactions.

Understanding how such eccentricities modify the wave launching process, and associated torque feedback on the perturber, is therefore important. In this work we will focus our attention on the low mass planetary regime where we can gain rigorous analytical insight by linearizing the fluid equations. The foundational work in this Type-I regime was laid out by \cite{GoldreichTremaine_1979,GoldreichTremaine_1980}. They argued that the wave excitation and coupling to the planetary potential is localised to Lindblad and corotation resonances. The total torque is then found by a summation over these contributions and only depends on the perturbing potential and disc properties at these discrete locations. Subsequent generalizations refined this picture by accounting for gradients in disc properties \citep[][]{Ward_1986} and by reintroducing non-WKB terms \citep[][]{Artymowicz_1993}. Nonetheless,  the aforementioned methods are fundamentally local in character. An alternative approach is to solve the full, linearized wave equation across the global disc without further approximation, as described by \cite{KorycanskyPollack_1993}. This method has proven successful in previous circular planet-disc studies, highlighting appreciable differences compared with the isolated resonant picture \citep[see][]{RafikovPetrovich_2012,Petrovich2012}, whilst also capturing subtle features in global torque density structures \citep[e.g.][]{MirandaRafikov_2019a,MirandaRafikov_2019b,CimermanEtAl_2024}. Furthermore, extended local approaches have shed light on three dimensional effects, whilst also accounting for small orbital eccentricities/inclinations, by means of numerical solutions to modified shearing box wave equations \citep[][]{TanakaEtAl_2002,TanakaWard_2004,TanakaOkada_2024}.

Inspired by these successes, this global methodology was recently extended to eccentric perturber orbits by \cite{FairbairnRafikov_2022} (hereafter \FRI), which found that the linear theory is capable of capturing the detailed, fine structure seen in the complex wake morphologies simulated by \cite{ZhuZhang_2022}. \cite{FairbairnRafikov_2025} (hereafter \FRII) subsequently employed this machinery, undertaking an expansive study of the torque excitation and associated back-reaction onto the planetary orbital elements. They comprehensively explored a suite of disc profiles and eccentricities up to $0.12$, finding a stark transition in behaviour when the eccentricity exceeds the local disc aspect ratio. They also improved on previous torque and eccentricity evolution prescriptions found in the literature, which were heuristically motivated and often limited in their range of validity \citep[e.g.][]{PapaloizouLarwood_2000, CresswellNelson_2006, IdaEtAl_2020}. 

In order to verify these new results, it is desirable to compare with fully nonlinear, hydrodynamical simulations. Whilst there exists a host of numerical simulations investigating the circular planet-disc interaction problem \citep[e.g.][]{DAngeloLubow_2010,DuffellMacFadyen_2012,KleyEtAl_2012,JimenezMasset_2017} there are fewer targeting the eccentric regime. Indeed, whilst \FRII loosely compare with the 2D numerical results of \cite{CresswellNelson_2006}, that early simulation was viscous, limited to a single disc profile, allowed the planet to migrate over time, and had a somewhat sparse eccentricity sampling. Subsequent experiments were undertaken by \cite{CresswellEtAl_2007,CresswellNelson_2008} with similar restrictions, although they also considered 3D discs and a number of different disc density profiles. Despite the challenges in quantitatively benchmarking the results of \FRII with these studies, there are encouraging features of qualitative overlap. In particular, the enhancement in the semi-major axis and eccentricity damping rates when the eccentricity is comparable to the disc aspect ratio was noted \citep[also see][]{BitschKley_2010}. More recently, \cite{PichierriEtAl_2023,PichierriEtAl_2024} performed 2D and 3D simulations of planet-disc interactions for low to intermediate eccentricity perturbers in a single disc profile. They focused on a number of eccentricities ranging from the circular regime to the critical case that $e$ exceeds the local aspect ratio, where they also found an enhancement in damping. 

These tantalising features demand a more targeted study which can be directly compared with the global, linear method of \FRII. Therefore, in this paper we will perform a tailored set of 2D simulations which examine planet-disc interaction across a range of disc profiles whilst pushing towards high eccentricity values of $\leq 0.7$. This significantly extends the previous upper limit of 0.12 explored in \FRII and challenges the capabilities of the analytical framework. With these bespoke simulations we will examine the wake morphology, radial excitation of torque and the net feedback on the planet, thereby verifying the linear theory in unprecedented detail. In section \ref{sec:linear_theory} we will review the methodology underpinning the linear results, before we describe the numerical setup in section \ref{sec:numerical_setup}. We will then present our results in section \ref{sec:results} and discuss them in section \ref{sec:discussion}. Finally we will conclude in section \ref{sec:conclusions}. Additionally, in appendix \ref{app:convergence} we discuss the convergence properties of our linear framework and numerical simulations, whilst appendix \ref{app:code_test} provides a repository of our data for benchmarking future numerical investigations.
% ================================%
\section{Summary of linear theory}
\label{sec:linear_theory}
% ================================%

In this section we will outline the governing equations for the linear, eccentric planet-disc interaction problem and briefly review the procedure by which they are solved. For more details the reader is referred to an in-depth description contained in \FRI, \FRII and references therein.

% ---------------------------- % 
\subsection{Basic approach}
\label{subsec:linear_approach}
% ---------------------------- %

We adopt a simple two-dimensional, coplanar and inviscid disc model about a star of mass $M_\star$, described in terms of the cylindrical coordinates $(r,\phi)$, similar to that employed by \cite{LinPapaloizou_1986}. The axisymmetric equilibrium disc is specified by the radial profiles of surface density $\Sigma$ and the vertically-integrated pressure $P = c_{\textrm{s}}^2 \Sigma$, where $c_{\rm s}$ is the isothermal sound speed. The equation for radial force balance then specifies the angular velocity $\Omega$, such that the azimuthal velocity $u_\phi = r\Omega$ while the radial velocity $u_r = 0$.

We now consider the linear perturbing effect of a planet with mass $M_{\rm p}$, which is inserted into the disc. This linear approach formally requires that $M_{\rm p} \ll M_{\rm th}$, where the thermal mass $M_{\rm th} = h_{\rm p}^3 M_\star$ and $h_{\rm p}$ denotes the disc aspect ratio at the planetary semi-major axis location. Specifically we set this low mass planet on a fixed Keplerian orbit with eccentricity $e$, semi-major axis $a_{\rm p}$ and resulting mean motion $n_{\rm p} = \sqrt{G M_\star/a_{\rm p}^3}$.\footnote{A schematic of this setup is visualised in Fig.1 of \FRI.} Throughout this work, we consider the planet-disc interaction in the astrocentric frame centred on the stellar primary and only include the \textit{direct} perturbing potential,
\begin{equation}
    \label{eqn:direct_potential}
     \Phi(r,\phi,t) = -\frac{GM_{\text{p}}}{\sqrt{r^2+R(t)^2-2 r R(t) \cos\left[\phi-\psi(t)\right]+\epsilon^2}}
\end{equation}
Here, $R(t)$ and $\psi(t)$ denote the time-dependent planetary coordinates whilst $\epsilon$ is a smoothing length, chosen to soften the singularity at the planet location according to a simple Plummer prescription. In this work, we have neglected the effect of the \textit{indirect} term arising from the non-inertial frame. Whilst including this term can lead to physically interesting and previously unappreciated features in the disc angular momentum flux \citep[e.g. Rafikov et al. in prep,][]{CridaEtAl_2025}, the dominant wake structure and torque feedback on the planet is not strongly affected (see \FRII). Since the orbit is non-circular, one cannot remove the fundamental time-dependence of the direct forcing by simply moving into a frame corotating with the planetary mean motion. Instead, the perturbing potential must be expanded in terms of two Fourier numbers -- in azimuthal space and orbital time -- such that,
\begin{equation}
    \label{eqn:potential_expansion}
    \Phi = \sum_{m = 1}^{\infty} \sum_{l = -\infty}^{\infty} \Phi_{ml}(r)\cos\left(m\phi-l n_{\rm p} t\right) .
\end{equation}
Here, the cosine dependence expresses the freedom to choose the phase of the orbit such that pericentre passage occurs at $t=0$ and the perturbing components $\Phi_{ml}$ are real. This external potential forces a linear response in the fluid variables. The net fluid quantities are therefore written as $X+\delta X$ where $X$ represents the equilibrium background whilst $\delta X$ is the small perturbation. These perturbations are similarly assumed to have a Fourier form,
\begin{equation}
    \label{eqn:modal_repsonse}
    \delta X_{ml} = \mathrm{Re}\left[\delta X_{ml}(r)e^{\mathrm{i}m(\phi-\omega_{ml}t)} \right],
\end{equation}
where $\omega_{ml} = (l/m) n_{\rm p}$ denotes the modal pattern speed (generally different from the planetary mean motion when $l \neq m$). Upon linearizing the governing continuity and Euler equations, then inserting the modal ansatz, one arrives at 
\begin{align}
\label{eq:density_pert_eqn}
&    -i \Tilde{\omega}\delta \Sigma + \frac{1}{r}\frac{d}{d r}(r\Sigma \delta u_r)+\frac{i m \Sigma}{r}\delta u_\phi = 0 , 
    \\
\label{eq:ur_eqn}
&    -i \Tilde{\omega}\delta u_r - 2\Omega \delta u_{\phi} = -\frac{1}{\Sigma}\frac{d}{d r} \delta P +\frac{1}{\Sigma^2}\frac{dP}{dr}\delta \Sigma-\frac{d}{d r}\Phi_{ml},
    \\
\label{eq:uphi_eqn}
& -i \Tilde{\omega}\delta u_\phi + \frac{\kappa^2}{2\Omega}\delta u_r = -\frac{i m}{r}\left( \frac{\delta P}{\Sigma}+\Phi_{ml}\right),
\end{align}
where $\kappa^2 = (2\Omega/r)d(r^2\Omega)/dr$ is the squared epicylic frequency and $\tilde{\omega} = m(\omega_{ml}-\Omega)$ is the Doppler shifted forcing frequency. Also note that here we have dropped the subscripts $ml$ for ease of notation. Equations \eqref{eq:density_pert_eqn}-\eqref{eq:uphi_eqn} are closed by specifying an equation of state (EoS). Throughout this paper we adopt the locally isothermal approximation for which the temperature $T_{\rm iso}(r) \propto c_s^2$ is a fixed function of radius and $\delta P = c_s^2 \delta \Sigma$.\footnote{The variable $T$ will represent a number of quantities throughout this paper (e.g. temperatures, torques and timescales), so particular care should be taken to note the distinguishing subscript notation.} We choose to focus on this simple thermodynamic model so as to directly interface with the comparable approach adopted in \FRI and \FRII. However, it should be remembered that alternative EoS prescriptions can lead to different wave behaviour -- as seen for adiabatic and `beta-cooling' discs by \cite{MirandaRafikov_2020}. Finally, combining equations \eqref{eq:density_pert_eqn}-\eqref{eq:uphi_eqn} in favour of the pseudo-enthalpy variable, $\delta h = \delta P/\Sigma$, eventually yields the master, modal, linear equation,
\begin{align}
\label{eq:master_eqn}
     \frac{d^2}{dr^2}\delta h_{ml} & +\left\{ \frac{d}{dr}\ln\left(\frac{r\Sigma}{D}\right)-\frac{1}{L_T} \right\} \frac{d}{dr}\delta h_{ml} \nonumber \\
    & - \left\{ \frac{2m\Omega}{r\Tilde{\omega}}\left[\frac{1}{L_T}+\frac{d}{dr}\ln\left(\frac{\Sigma \Omega}{D}\right)\right] \right. \nonumber \\
    & \left.+\frac{1}{L_T}\frac{d}{dr}\ln\left(\frac{r\Sigma}{L_T D}\right)+\frac{m^2}{r^2}+\frac{D}{c_{\textrm{s}^2}}\right\} \delta h_{ml} = \nonumber \\
    & = -\frac{d^2 \Phi_{ml}}{dr^2} - \left[ \frac{d}{dr}\ln\left(\frac{r\Sigma}{D}\right)\right]\frac{d\Phi_{ml}}{dr} \nonumber \\
    & + \left\{\frac{2m\Omega}{r\Tilde{\omega}}\left[\frac{d}{dr}\ln\left(\frac{\Sigma \Omega}{D}\right)\right]+\frac{m^2}{r^2}\right\}\Phi_{ml},
\end{align}
which should be solved for each combination of $(m,l)$. Here $L_\textrm{T}$ denotes the characteristic locally isothermal length scale set by
\begin{equation}
    \label{eqn:thermal_scale}
    \frac{1}{L_\textrm{T}} = \frac{d \ln c_\textrm{s}^2}{dr} ,
\end{equation}
whilst $D$ measures the offset from a Lindblad resonance given by
\begin{equation}
    \label{eqn:resonance_offset}
    D = \kappa^2-\tilde{\omega}^2.
\end{equation}

% ---------------------------- % 
\subsection{Disc structure}
\label{subsec:disc_structure}
% ---------------------------- %

In order to make things definite, we must assume a particular equilibrium, unperturbed disc structure. Throughout this paper, we will adopt smooth power-law profiles such that the surface density is given by
\begin{equation}
    \label{eqn:surface_density}
    \Sigma(r) = \Sigma_{\rm p}\left(\frac{r}{a_{\rm p}}\right)^{-p} ,
\end{equation}
and the locally isothermal sound speed,
\begin{equation}
    \label{eqn:sound_speed}
    c_\textrm{s} (r) = c_{\rm s,p}\left(\frac{r}{a_{\rm p}}\right)^{-q/2},
\end{equation}
which yields a temperature profile, $T_{\rm iso} \propto c_s^2 \propto r^{-q}$. Here and henceforth, all the disc quantities subscripted by ‘p’ are characteristic values evaluated at the planetary semi-major axis. Assuming vertical hydrostatic balance, the aspect ratio of the disc follows 
\begin{equation}
    \label{eqn:aspect_ratio}
    h(r) \equiv \frac{H(r)}{r} = h_\textrm{p} \left(\frac{r}{a_{\rm p}}\right)^{(1-q)/2},
\end{equation}
where $H(r) = c_s(r)/\Omega_\textrm{K}(r) $ is the pressure scale height and $\Omega_\textrm{K}(r) = \sqrt{GM_{*}/r^3}$ denotes the local Keplerian velocity. Therefore, we can identify $c_{\rm s, p} = h_{\rm p} a_{\rm p} n_{\rm p}$.

% ---------------------------- % 
\subsection{Solving the equations}
\label{subsec:solving_linear}
% ---------------------------- %

Having derived the master equation \eqref{eq:master_eqn}, encapsulating the excitation of the density wake, we will now briefly summarize our solution method. Whilst purely analytical techniques, such as that pioneered by \cite{GoldreichTremaine_1979,GoldreichTremaine_1980}, typically consider the local wave excitation process in the vicinity of resonance locations, we instead numerically solve for the global structure between $r_{\rm in} = 0.05 a_{\rm p}$ and $r_{\rm out} = 5 a_{\rm p}$. This approach was first described by \cite{KorycanskyPollack_1993} and proved successful in subsequent circular planet-disc interaction studies \citep[see][]{MirandaRafikov_2019a,MirandaRafikov_2020} where the same master equation manifested. More recently, this technique has been extended towards the eccentric planetary regime in \FRI and \FRII, where the essential difference is that the Fourier decomposition of the perturbing potential depends on two quantum numbers, $(m,l)$. Therefore, one must compute the wave response for many more forcing modes. Intuitively, $m$ governs the spatial detail of the wake and thus we will require many modes to sufficiently capture the fine structure. Meanwhile, $|m-l|$ measures departures from a circular orbit, for which $m=l$. Hence the more eccentric orbits, and thus more time-dependent wakes, demand higher `off-diagonal' $l$ contributions. In practice, we extract each $\Phi_{ml}$ contribution via a numerical double integral across a grid of radii, using a trapezoidal quadrature with 1024 points in both $\phi$ and $t$. We then interpolate onto arbitrary $r$ as required by the ordinary differential equation (ODE) solver. The master equation is integrated out from corotation resonances, where $\omega_{ml} = \Omega$, and outgoing radiative boundary conditions are applied to fix the contributions from the homogeneous and forced wave components. Additional corrections, such as the phase-gradient error minimisation procedure, slightly adjust the boundary conditions in order to reduce contamination from inwardly propagating wave components, effectively maximising the transmission coefficients of outgoing waves \citep[for details see][]{KorycanskyPollack_1993}.

% ---------------------------- % 
\subsection{Important diagnostics}
\label{subsec:linear_diagnostics}
% ---------------------------- %

The enthalpy modes can be converted into more useful quantities such as $\delta\Sigma_{ml}$, $\delta u_{r,ml}$ and $\delta u_{\phi,ml}$ via equations \eqref{eq:density_pert_eqn}-\eqref{eq:uphi_eqn}. We can then sum these across all $(m,l)$ contributions to recover net perturbation maps for the density $\delta\Sigma$, radial velocity $\delta u_r$ and azimuthal velocity $\delta u_\phi$ (see section \ref{subsec:wake_morphology}). These linear wave modes also allow us to compute second-order quantities such as the torque density and angular momentum flux (AMF) which govern the planet-disc coupling. Indeed, the AMF is defined as 
\begin{equation}
    F_{J}(r)  = r^2 \Sigma(r) \oint \operatorname{Re}[\delta u_r(r,\phi)] \operatorname{Re}[\delta u_\phi (r,\phi)] d\phi ,
    \label{eq:AMF-def}
\end{equation}
whilst the radial torque density is
\begin{equation}
    \label{eq:dTdr_def}
    \frac{dT}{dr} = -r \oint \operatorname{Re}[\delta\Sigma(r,\phi)]\operatorname{Re}\left[\frac{\partial \Phi}{\partial\phi}\right]d\phi.
\end{equation}
These quantities can be computed at any instant from the net perturbation maps. Alternatively, pursuing a more elegant approach, one may insert the modal expansion for each perturbed variable, given by equation \eqref{eqn:modal_repsonse}, into equations \eqref{eq:AMF-def}-\eqref{eq:dTdr_def} and then exploit integral orthogonality conditions to simplify matters (see \FRII). Furthermore, we are interested in the orbit-averaged quantities enacted through the operator $\langle\cdot\rangle = (1/T_{\rm p})\oint\cdot dt$, where $T_{\rm p} = 2\pi/n_{\rm p}$ is the orbital period of the perturber. This averaging simplifies the expression further and we arrive at equations which encapsulate the individual modal contributions to the AMF and torque density,
\begin{equation}
    \label{eq:orbit_avg_amf_dtdr}
    \langle F_{J}(r) \rangle = \sum_{m,l} F_{J, ml} 
    \quad \textrm{and} \quad
    \left\langle dT/dr \right\rangle = \sum_{m,l} dT_{ml}/dr ,
\end{equation}
where
\begin{align}
    \label{eq:F_jml}
    & F_{J, ml} = \pi r^2 \Sigma(r) \operatorname{Re}[\delta u_{r, ml} \delta u_{\phi, ml}^{*}], \\
    \label{eq:dTdr_jml}
    & dT_{ml}/dr = -\pi r m \Phi_{ml} \operatorname{Im} [\delta \Sigma_{ml}].
\end{align}
Henceforth, we will discard the angled brackets and simply bear in mind that we are dealing with the time-averaged results. Integrating these torque densities across the full radial extent and adding the contributions from all modes yields the net torque $T$ enacted on the entire disc. Additionally, the conservation of the planet's Jacobi integral under the gravitational back-reaction of each uniformly rotating wave mode, relates the planetary orbital energy $E$ to the torque and pattern speed \citep[see][]{FairbairnRafikov_2025}. Indeed one can show that 
\begin{equation}
    \label{eq:jacobi_energy}
    \tau_{\rm E}^{-1} \equiv \frac{1}{E}\frac{dE}{dt} = (2/{M_{\rm p}a_{\rm p}^2 n_{\rm p}^2})\sum_{m,l} \omega_{ml}T_{ml}.
\end{equation}
This exchange of angular momentum and energy will ultimately drive the evolution of the planetary orbital elements. The angular momentum of a Keplerian orbit is 
\begin{equation}
\label{eq:AM_planet}
    L = M_{\rm p}\sqrt{GM_\star a_{\rm p} (1-e^2)} = M_{\rm p} n_{\rm p} a_{\rm p}^2 \sqrt{1-e^2} ,
\end{equation}
whilst the orbital energy is
\begin{equation}
    \label{eq:E_planet}
    E = -\frac{GM_\star M_{\rm p}}{2 a_{\rm p}} = -\frac{1}{2} M_{\rm p} n_{\rm p}^2 a_{\rm p}^2.
\end{equation}
Differentiating these with respect to time \citep[e.g.][]{IdaEtAl_2020, AtaieeKley_2021}, yields
\begin{equation}
\label{eq:timescale_relation}
    \tau_{L}^{-1} = \frac{1}{2}\tau_{a}^{-1}-\frac{e^2}{1-e^2}\tau_{e}^{-1} \quad \rm{and} \quad \tau_{\rm E}^{-1} = \tau_{a}^{-1},
\end{equation}
where the characteristic evolution rates are given by
\begin{equation}
\label{eq:inverse_timescales}
    \tau_{L}^{-1} = -\frac{1}{L}\frac{dL}{dt} , \quad \tau_{a}^{-1} = -\frac{1}{a_{\rm p}}\frac{da_{\rm p}}{dt} , \quad \tau_{e}^{-1} = -\frac{1}{e}\frac{de}{dt}.
\end{equation}
Due to the equal and opposite exchange of angular momentum between the disc and planet, we can immediately identify $\tau_{L}^{-1} = T/L$.  Meanwhile, the relationship between energy and semi-major axis allows us to identify $\tau_a^{-1}$ with the characteristic energy rate given by $\tau_{\rm E}^{-1}$. Finally, $\tau_{e}^{-1}$ is trivially obtained by rearrangement of the first equation in \eqref{eq:timescale_relation}.

% ---------------------------- % 
\subsection{Solution parameters}
\label{subsec:linear_parameters}
% ---------------------------- %

Equipped with the linear theory and solution methodology summarized above, we are now in a position to outline the target parameter space. Whilst \FRII comprehensively explores a grid of $(q,p,h,e)$ values, here we will be more selective since our emphasis will be placed on detailed comparisons with a discerning number of fully nonlinear, numerical experiments (as described in section \ref{sec:numerical_setup}) which push to higher eccentricities than previously studied. To this end, we will adopt three fiducial disc profiles with $(p,q) \in \{(1.5,1.0),(1.5,0.0),(0.5,0.0)\}$. In each case, we will compute the wake response for a wide range of $e$ ranging between 0.0 and 0.7. For all calculations, we adopt a reference scale height $h_{\rm p} = 0.06$ and a smoothing length $\epsilon = 0.3 h_{\rm p} a_{\rm p}$. Each calculation will adopt a default modal range of $m \in (1,170)$ and $|l-m| \in (0,40)$, which proves to be sufficient for attaining good convergence with the numerical experiments. We address the nature of this convergence with varying $|l-m|$ in appendix \ref{app:convergence}. We are also careful to exclude contributions from certain combinations of $(m,l)$ in our final modal summation. In particular, we ignore modes for which the corresponding pattern speeds resonate with Lindblad locations lying exterior to our solution domain. Otherwise our linear solver tries to impose outgoing radiative boundary conditions when in fact there is no wave excited within the domain, potentially leading to spurious results.
% ================================%
\section{Numerical setup}
\label{sec:numerical_setup}
% ================================%
For comparison with the linear theoretic calculations described above, our hydrodynamical simulations solve the equations of two-dimensional (vertically-integrated) inviscid hydrodynamics. The continuity and momentum equations are given by
\begin{equation}\label{eq:continuity}
\partial_t\Sigma + \nabla\cdot(\Sigma \mathbf{v})=0
\end{equation}
and
\begin{equation}\label{eq:momentum}
\partial_t(\Sigma \mathbf{v}) + \nabla\cdot(\Sigma\mathbf{v}\mathbf{v}+P\mathbb{I})=-\Sigma\nabla\Phi_T,\\
\end{equation}
respectively, where where $\mathbf{v}$ is the fluid velocity, $\mathbb{I}$ is the identity tensor, and $\Phi_T$ is the full gravitational potential. As in the preceding linear theory, we take the vertically-integrated pressure to be $P=c_s^2(r)\Sigma$. 

We approximate solutions to Eqns. \ref{eq:continuity}---\ref{eq:momentum} using the moving-mesh finite volume code \Disco \citep{2016ApJS..226....2D}.\footnote{
Specifically, we used the version \url{https://github.com/ajdittmann/Disco}, which includes additional optimizations and in-situ diagnostics.} To achieve very low numerical dissipation when simulating thin (thus having a high azimuthal Mach number) accretion disks, we adopted in \Disco a moving-mesh strategy in which each grid annulus rotated with the average angular velocity of the fluid in that annulus. The simulations used second-order Runge-Kutta time stepping \citep{1998MaCom..67...73G} and a second-order total-variation-diminishing spatial scheme \citep{{1979JCoPh..32..101V},{2000JCoPh.160..241K}} along with a Harten-Lax-Van Leer-Contact approximate Riemann solver \citep{1994ShWav...4...25T}.

Our fiducial simulations used $N_r=4608$ cells in the radial direction, which were linearly spaced from $r=0.05r_p$ to $r=0.1r_p$ and geometrically spaced from $r=0.1r_p$ to $4.5r_p$. The number of cells in each annulus was chosen to maintain cell aspect ratios as close to unity as possible, typically resulting in $N_\phi\approx 6528$. We employed damping zones near the inner and outer boundaries $r\in[0.05,0.1]\cup[4.0,4.5]$ which damped the conserved variables to their values at $t=0$ to prevent spurious wave reflections.
This discretization resolved the scale height of the disk at $r=a$ by 64 cells, which yielded converged torque measurements and orbital evolution results (see appendix \ref{subsec:simulation_convergence}) but was insufficient to resolve the weakly nonlinear damping of the excited spiral arms (see section \ref{subsec:amf_deposit}).

To make direct comparison with the linear calculations, we centre the cylindrical coordinate system of the simulations on the star but neglect the non-inertial forces that should be introduced in such a frame, equivalent to the neglect of the indirect potential in section \ref{sec:linear_theory}. For more details on the consistency of this approach, see the recent work by \cite{CridaEtAl_2025} and Rafikov et al., in prep. Thus, the full gravitational potential is given by that of the star in addition to the direct planetary potential introduced in Eqn. \ref{eqn:direct_potential}: $\Phi_T = -M_\star/r + \Phi$.

Our simulations also record various azimuth- and area-averaged quantities, which are also time-averaged along with the Runge-Kutta timestepping. The most important of these are the torque and power delivered to the planet during its interaction with the disk, 
\begin{align}
\left.\frac{dL}{dt}\right|_i &= \frac{1}{t_i-t_{i-1}}\int_{t_{i-1}}^{t_i}dt\int \Sigma\mathbf{R}\times\nabla \Phi_T dA \\
\left.\frac{dE}{dt}\right|_i &= \frac{1}{t_i-t_{i-1}}\int_{t_{i-1}}^{t_i}dt\int \Sigma \mathbf{v}\cdot\nabla \Phi_T dA,
\end{align}
where $\mathbf{R}$ is the position vector of the planet and $\mathbf{v}$ is its velocity. 
We calculate time-averaged torque density and (wave-driven) angular momentum flux similarly. Defining $\bar{\Sigma}(r)$ as the azimuth and time-averaged surface density and $\langle...\rangle$ is now a mass-weighted time and azimuth average, the average torque density excited in the disk is $dT/dr = 2\pi r \bar{\Sigma}\langle \mathbf{r}_p\times\nabla\Phi_T\rangle$, and the (wave) angular momentum flux $F_J=2\pi r^2\bar{\Sigma}(\langle v_r v_\phi\rangle - \langle v_r\rangle\langle v_\phi\rangle)$.

Our simulations neglect disc self-gravity and are thus largely scale-free (the surface density can be rescaled arbitrarily and we freely chose time units such that $G(M_p+M_\star)=1$). However, unlike the linear theory presented earlier the mass of the secondary necessarily enters the governing equations and controls the strength of various nonlinearities. Apart from a limited number of explorations in section \ref{subsubsec:higher_mass}, we set the planetary mass ratio $q_r=M_p/M_\star=10^{-6}$, and thus $M_p=0.0046 M_{\rm th}$, well within the linear regime. Our fiducial simulations are then run for 25 planetary orbits, which is sufficient time for establishing the linear wake response, whilst not being so long as to start saturating the corotation torque. A selection of longer runs, lasting 200 planetary orbits, will be performed in section \ref{subsubsec:corotation} in order to probe this nonlinear saturation effect. 
% ================================%
\section{Results}
\label{sec:results}
% ================================%

In this section we will compare the results from linear theory, developed in section \ref{sec:linear_theory}, with the fully nonlinear, hydrodynamical experiments described in section \ref{sec:numerical_setup}. To this end, we will address a number of key diagnostics, each of which is examined in detail in the following subsections. In summary, we find exquisite  qualitative and quantitative agreement between the methodologies for a wide range of disc properties and planetary eccentricities, lending complementary confidence in both techniques. 

%---------------------------------%
\subsection{Wake morphology}
\label{subsec:wake_morphology}
%---------------------------------%

As demonstrated by \FRI, the linear theory is capable of capturing the fine structure associated with the complex wake morphology found in previous simulations. They qualitatively compared their linear results with the numerical calculations of \cite{ZhuZhang_2022} for a single disc profile and claimed very good agreement. Furthermore, the separate studies were bridged by the efficacy of a simplified wavelet model, which tracks an azimuthally sheared radially propagating packet and captures the dominant eccentric wake features in both the simulation and the theory. Here, our own customised simulations offer a direct comparison across a wider range of disc parameters. 

\begin{figure*}
    \centering
    \includegraphics[width=\linewidth]{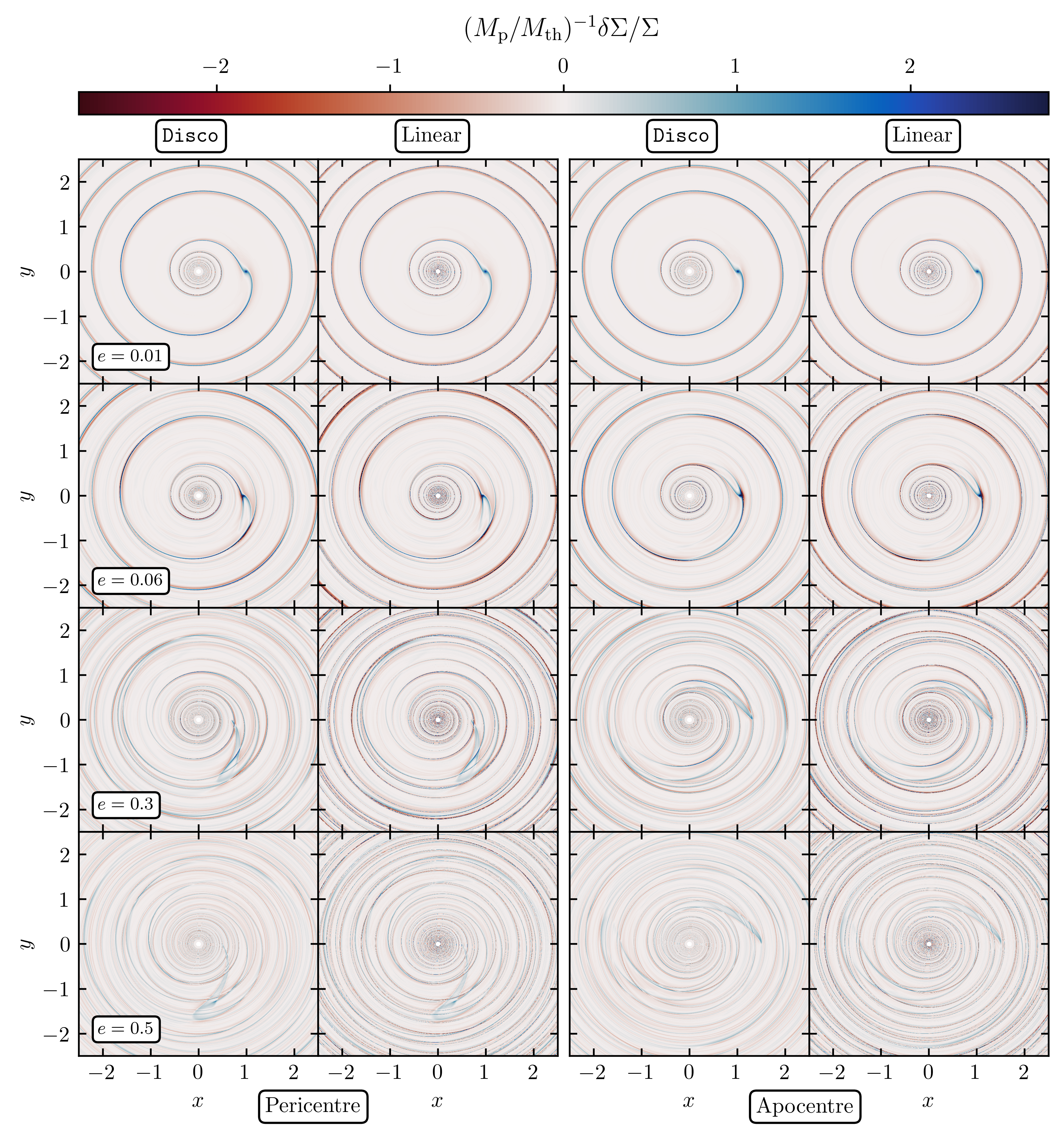}
    \caption{Maps of the relative surface density perturbation for the disc profile $(p,q) =(1.5,0.0)$. The rows denote increasing values of $e \in \lbrace 0.01,0.06,0.3,0.5 \rbrace$. The left two vs right two columns correspond to peri vs apocentre snapshots respectively. Within these, the first column shows the \texttt{Disco} results whilst the second shows the predictions of linear theory.}
    \label{fig:sigma_grid}
\end{figure*}

In Fig.~\ref{fig:sigma_grid} we plot the relative surface density perturbation for the specific case $(p,q) =(1.5,0.0)$. The rows show the results across a range of eccentricities $e \in \lbrace 0.01,0.06,0.3,0.5\rbrace$. The left two columns correspond to the pericentre phase whilst the right two columns show the planet at apocentre. These are further divided into the \texttt{Disco} and linear theory results. In all panels we immediately observe exquisite  morphological agreement between the simulations and theory -- extending from the near circular regime all the way to large eccentricities. For $e=0.01$ we see that the wake closely resembles the standard circular picture with an inwards and outwards propagating spiral arm and minimal differences between peri and apocentre. More notable structures emerge when $e\geq h_{\rm p}=0.06$ for which the radial velocity of the planet relative to the gas becomes transonic. Here we see that the planet can detach from the inner and outer wake structure as the epicylic motion exceeds the communication speed set by sound waves. If one looks very closely at the outer spiral wake for these $e = 0.01$ and $e = 0.06$ panels, it appears that the linear theory captures a slightly more concentrated arm with a higher amplitude. This can be attributed to the weak amount of nonlinear steepening and subsequent damping which occurs in the numerical simulations (see \cite{CimermanRafikov_2021} and section \ref{subsec:amf_deposit}).

Pushing towards higher eccentricities leads to more intricate patterns, although close examination still reveals impressive overlap between the theory and simulations for all the main features. In particular the supersonic motion leads to a wedge-like wake trailing the planet. Near pericenter the fast azimuthal motion counteracts the differential shear so the wake lags behind the planet, in contrast to the circular regime where the inner arm is pulled forwards. Due to the time-variability of the waves launched over the epicylic period, and the dilution of wave action across many more modal components, there is less constructive interference and the maximal amplitudes decrease. 

\begin{figure*}
    \centering
    \includegraphics[width=\linewidth]{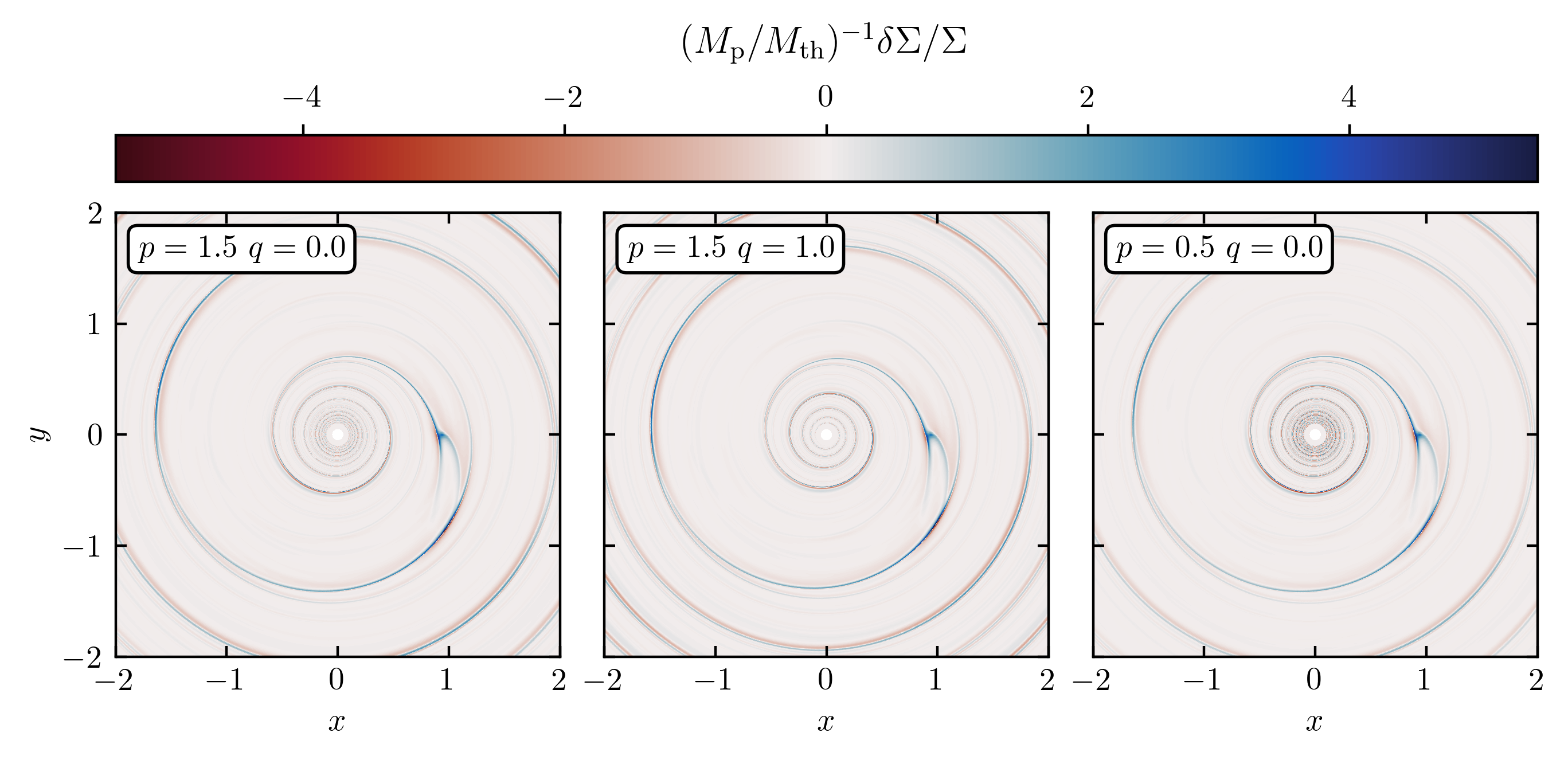}
    \caption{Maps of the relative surface density perturbation at pericentre from our \texttt{Disco} simulations for $e = 0.06$. The three columns show the disc profiles  $(p,q)$ = $(1.5,0.0)$, $(1.5,1.0)$ and $(0.5,0.0)$.}
    \label{fig:sigma_grid_exps}
\end{figure*}

To complement these plots, in Fig.~\ref{fig:sigma_grid_exps} we also examine the surface density maps from our simulation runs across the three disc profiles being investigated. For illustration, we adopt the critical value of $e = h_{\rm p} = 0.06$ and plot the wake at the pericentre snapshot. The main morphological features are very similar across all disc profiles with only subtle differences. Most notably, in the case with $q=1.0$ the sound speed decreases outwards so the exterior spiral has a smaller pitch angle and is more tightly wound whilst the interior wake is less tightly wound, compared with the globally isothermal runs.

\begin{figure}
    \centering
    \includegraphics[width=\linewidth]{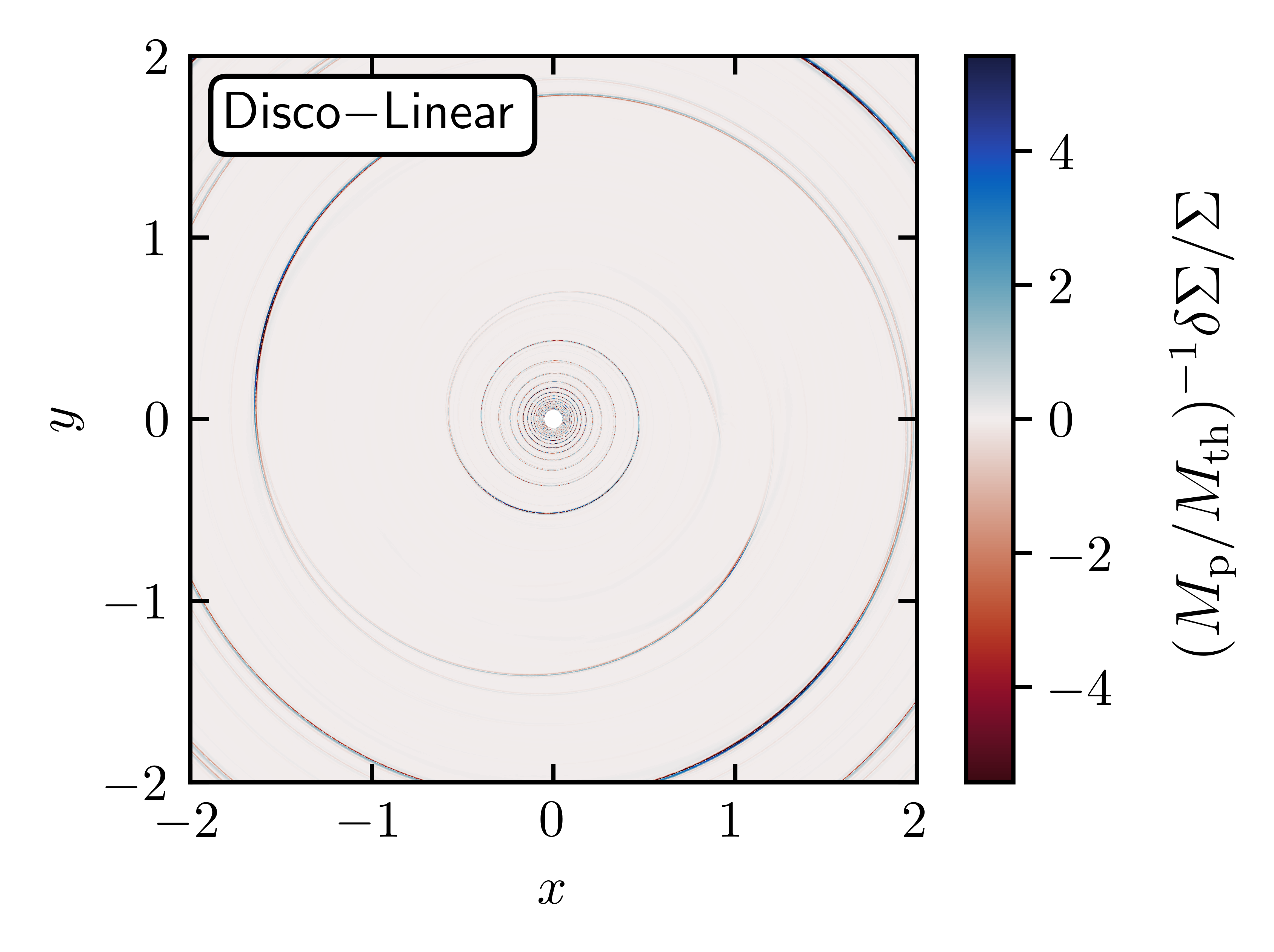}
    \caption{Surface density difference map comparing the linear theory and fiducial simulation results for the disc with $(p,q) = (1.5,0.0)$ and $e=0.06$.}
    \label{fig:sigma_diff}
\end{figure}

It should be acknowledged that the comparison thus far has been fairly qualitative. Whilst one might be tempted to produce difference maps between the simulation and theory results to quantify the discrepancies, these are complicated by the combination of weakly nonlinear, shock-induced damping of the wave and numerical dissipation of the wake structure (discussed further in section \ref{subsec:amf_deposit}). Furthermore, slight shifts in the very narrow wake structures can lead to order unity differences. Indeed, in Fig.~\ref{fig:sigma_diff} we demonstrate this by plotting the residual surface density differences between the simulations and linear theory for the particular disc profile with $(p,q) = (1.5,0.0)$ and $e = 0.06$. We see that there is good agreement in the vicinity of the planet and its coorbital radius, although further away there are clearly growing residuals. Despite these residuals, it should be emphasised that the qualitative wake structure is very good. In the upcoming sections we will find that the time/azimuth-averaged and radially integrated diagnostics allow for a more meaningful quantitative comparison. 

%---------------------------------%
\subsection{Torque profiles}
\label{subsec:torque_profiles}
%---------------------------------%

\begin{figure*}
\includegraphics[width=\linewidth]{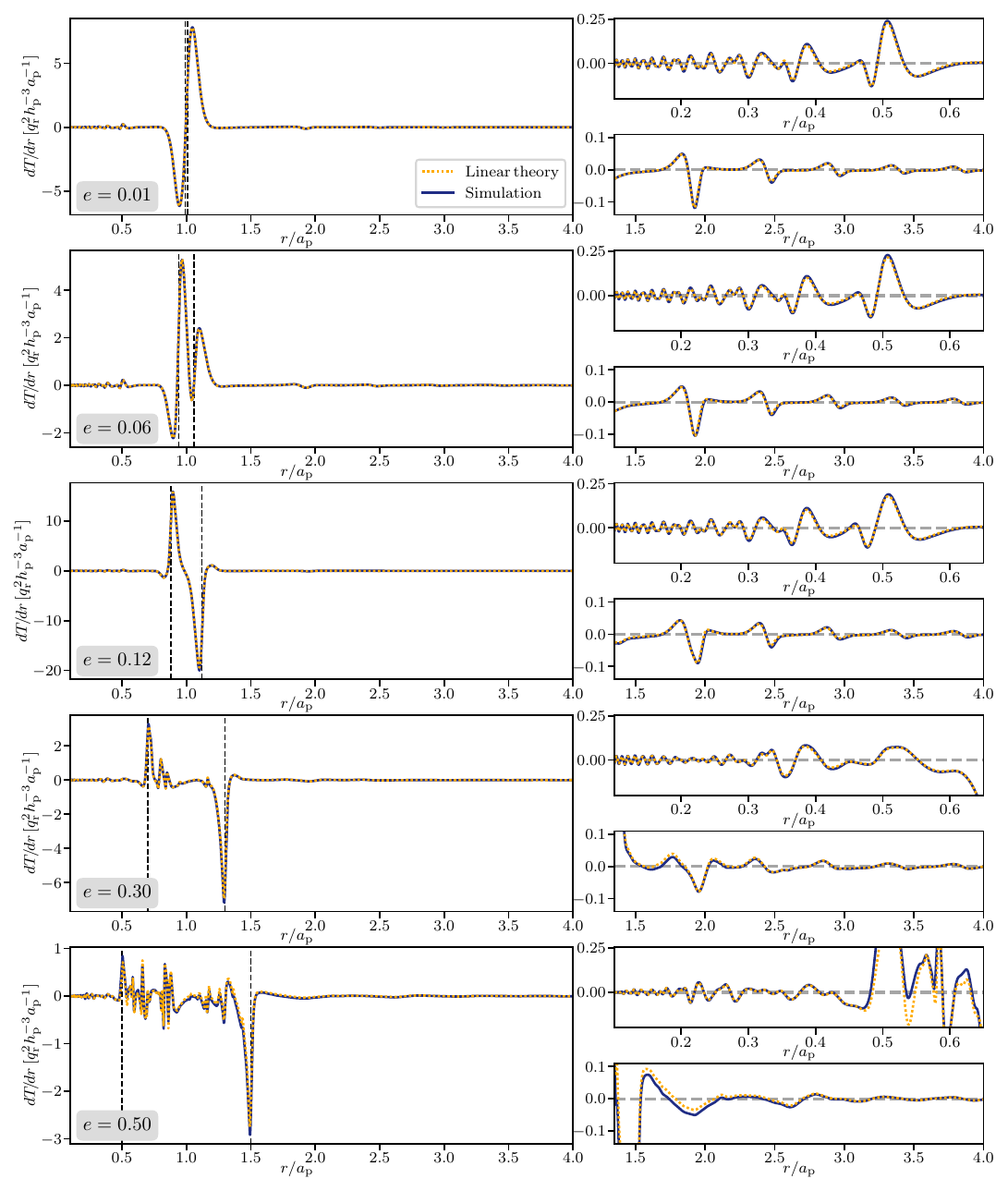}
\caption{Comparison of the radial torque density profiles for the disc parameters $(p,q)=(1.5,0.0)$ obtained using linear theory (\textit{yellow dashed}) and simulations (\textit{blue solid}). \textit{Left column}: shows the full radial $dT/dr$ structure for a range of eccentricities $e \in \{0.01,0.06,0.12,0.3,0.5\}$ moving down the rows. The vertical dashed lines denote the pericentre and apocentre radial extremities of the planetary orbit. \textit{Right column}: shows the corresponding zoomed in torque wiggles in the inner and outer disc.}
\label{fig:torque_p1.5q0.0}
\end{figure*}

\begin{figure*}
\includegraphics[width=\linewidth]{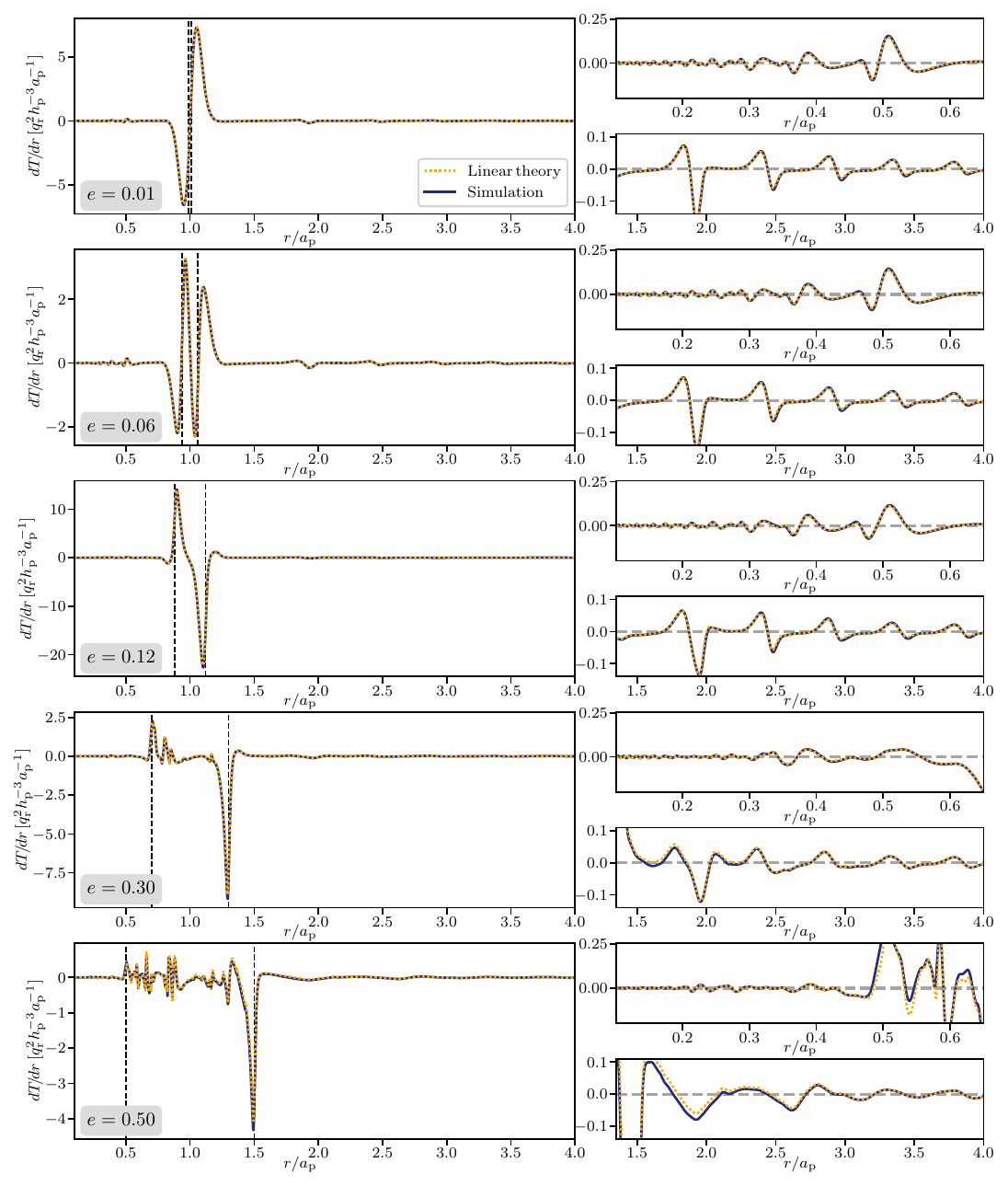}
\caption{The same as Fig.~\ref{fig:torque_p1.5q0.0} but for the disc profile $(p,q)=(0.5,0.0)$.}
\label{fig:torque_p0.5q0.0}
\end{figure*}

We now examine in detail the orbit-averaged radial torque density profiles, defined according to equations \eqref{eq:dTdr_def}-\eqref{eq:orbit_avg_amf_dtdr}, for a number of slices through our data. Figs.~\ref{fig:torque_p1.5q0.0} and \ref{fig:torque_p0.5q0.0} show the cases with $(p,q)$ equal to $(1.5,0.0)$ and $(0.5,0.0)$, respectively. In both examples the left-hand column displays the torque density profile across the full numerical domain, whilst the right-hand column zooms in to the fine structure in the inner and outer disc. The results are plotted in characteristic units of $F_{J,0}/a_{\rm p}$ where $F_{J,0} \equiv \Sigma_{\rm p} a_{\rm p}^4 n_{\rm p}^2 h_{\rm p}^{-3} (M_{\rm p}/M_\star)^2$ is the typical angular momentum exchanged by the interaction. For our choice of code units this simply becomes $F_{J,0} = q_{\rm r}^2/h_{\rm p}^3$. Each row moving downwards displays the results for increasing values of $e \in \{0.01,0.06,0.12,0.30,0.50\}$, spanning the near-circular regime, via the transonic eccentricity threshold $e = h_{\rm p}$, towards highly eccentric orbits. The solid blue lines denote the simulation results, whilst the dotted yellow line shows the comparison with linear theory. The vertical dashed lines indicate the planetary pericentre and apocentre locations. In all figures, we immediately note the exquisite  overlap between the simulations and linear theory, even for the highest values of eccentricity. This simultaneously suggests that the numerical setup is very well behaved and also that the linear theory has included sufficient modes for convergence (see appendix \ref{app:convergence}). 

% ...............................%
\subsubsection{Behaviour with $e$}
% ...............................%

Presently, let us focus our attention on Fig.~\ref{fig:torque_p1.5q0.0} for the simplest disc structure with $(p,q)=(1.5,0.0)$, for which the disc is globally isothermal and has negligible vortensity gradients, thus precluding potentially worrisome corotation torques \citep[][]{GoldreichTremaine_1979}. For $e=0.01$ we see a standard torque profile, close to that of the typical circular planet-disc interaction. The main torque contributions are composed of a peak and a trough on either side of the planetary semi-major axis. These are sourced by dominant Lindblad resonances in the vicinity of $|r-a_{\rm p}| \sim H_{\rm p}$. Closer to the planet, the `torque-cutoff' phenomenon exponentially suppresses the excited torque from high-order $m$ contributions \citep[e.g.][]{GoldreichTremaine_1980,Artymowicz_1993}. 

As higher values of $e$ are considered, the torque density changes notably. Specifically, when $e = 0.06 = h_{\rm p}$ the planet performs transonic radial motion relative to the gas, drastically altering the wake structure and associated torques. Similarly to panel (a) in Fig.~5 of \FRII (for $p=q=0.0$), we see that the peaks and troughs develop more radial oscillations in the vicinity of the planet. Pushing into the supersonic regime with $e=0.12$ (the highest value previously considered in \FRII) the torque peak and trough ordering has now reversed compared with the circular regime. The location of these maximum and minimum turning points coincide with the apocentre and pericentre of the eccentric planetary orbit (marked by the vertical dashed lines), indicating that most of the angular momentum exchange is facilitated during these orbital phases. As intuitively explained by \cite{PapaloizouLarwood_2000} and \cite{MutoEtAl_2011}, the planet trails/leads the excited wake at apo/pericentre and hence pulls on the gas in the opposite sense to the standard circular planet-disc interaction -- thus explaining the reversal in the torque profile. Furthermore, we have previously seen in Fig.~\ref{fig:sigma_grid_exps} that the wake structure in $\delta\Sigma/\Sigma$ is essentially insensitive to the surface density profile, $p$. Therefore, with all else approximately equal, as the surface density becomes shallower, the negative torque contribution on the disc exerted near apocentre becomes stronger whilst the positive torque around pericentre becomes weaker, hence driving a deeper torque reversal (compare Figs.~\ref{fig:torque_p1.5q0.0} and \ref{fig:torque_p0.5q0.0} for $e \geq 0.12$). Additionally, for low $p$ and high $e$, the positive power enacted on the planet at apocentre can eventually dominate and promote outwards orbital migration as we will see later in section \ref{subsec:orbital_evolution}.

Pushing towards highly supersonic orbits with $e = 0.3$, the extremal turning points are wider still as the peri/apocentre are more separated. For these very eccentric orbits, the increased time spent at apocentre strongly weights the interaction there, leading to the growing dominance of the outer trough. In between these extrema, we observe that the torque profile becomes much more jagged, as one might anticipate from the highly complex wake structures observed in section \ref{subsec:wake_morphology}. Nonetheless, the overlap between the linear theory and simulations is still remarkable throughout this region, with only slight discrepancies discernible in the peak and trough amplitudes at the orbital limits. This complex behaviour becomes even more pronounced in the lowest row where we see the results for $e = 0.5$. Here the torque density exhibits rapid radial variations although there still seems to be decent agreement with the main features which will contribute most to the net torque. Indeed, whilst the addition of more linear modes may be able to better capture some of this fine structure, we find that the integrated quantities are already rather well converged for our fiducial modal range (see convergence study in appendix \ref{app:convergence}).

Fig.~\ref{fig:torque_p0.5q0.0}, which displays results for $p = 0.5$ and $q = 0.0$, reinforces the above findings. It is reassuring to see that we are able to attain remarkable agreement with the linear theory across more general disc parameters. The qualitative features are similar, although the magnitude of the torques and the detailed location of some of the peaks have shifted. One might be surprised at the continued good agreement given the fact that corotation torques enter the problem for such general disc profiles. In the vicinity of the planet the interaction becomes intrinsically nonlinear which can change the character of the corotation torque as fluid undergoes `horseshoe' orbits which exchange angular momentum with the planet \citep[e.g.][]{Ward_1991,PaardekooperPapaloizou_2009a}. However, given the low value of $q_{\rm r}$ it takes a long time for the nonlinear horseshoe effects to manifest and the classical, linear corotation torque is largely preserved at the time of measurement in the simulations (25 orbits). We will discuss the late time behaviour and emergence of the horseshoe drag further in section \ref{subsubsec:corotation}.

% ...............................%
\subsubsection{Torque wiggles}
% ...............................%

Whilst the torque density is almost reduced to zero in the far field away from the planet, the right hand column panels of Figs.~\ref{fig:torque_p1.5q0.0} and \ref{fig:torque_p0.5q0.0} zoom in to reveal the presence of `torque wiggles'. Once again we see remarkable agreement between the numerics and theory emphasising the ability of our simulations to capture subtle features. This phenomenon has previously been observed in circular planet simulations performed by \cite{ArzamasskiyEtAl_2018} and \cite{DempseyEtAl_2020}, before \cite{CimermanEtAl_2024} provided a detailed analytical understanding. In summary, they arise due to the coupling between the perturbing potential and the net density wave, as it crosses the line intersecting the star and the planet. This relies on the approximately self-similar structure of the density wake and is dominated by the low order $m$ modes as one move radially outwards from the planet.

Whilst this theory was originally understood in the context of the circular planet-disc interaction problem, here we note that the torque wiggle structure for the transonic $e = 0.06$ and $0.12$ runs is almost identical to the near-circular $e=0.01$ case. Indeed, far enough away from the planet, the epicylic motion becomes negligible and therefore the diagonal potential components $\Phi_{mm}$ will dominate the planetary coupling to the disc surface density. To gain further insight, it is therefore useful to compare the diagonal Fourier component structure of the surface density at two locations in the outer disc, following the analysis pipeline of Fig.~5 in \cite{CimermanEtAl_2024}.
\begin{figure}
\includegraphics[width=\linewidth]{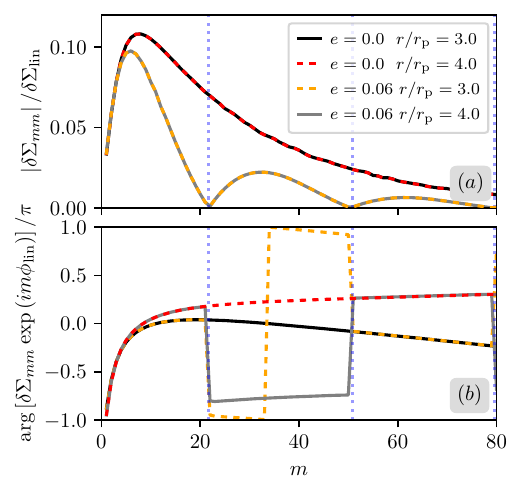}
\caption{Diagonal surface density modes $\delta\Sigma_{mm}$ governing the shape of the torque wiggles for the case $(p,q) = (1.5,0.0)$. \textit{Panel (a)}: the Fourier amplitudes for the surface density perturbations normalized by the scaling from the conserved wave action. \textit{Panel (b)}: the phase angle of the modes relative to the dominant spiral arm phase predicted from linear theory. The curves correspond to two values of $e$ at two representative radii. $(e,r/a_{\rm p}):$ (0.0,3.0) = black solid, (0.0,4.0) = red dashed, (0.06,3.0) = orange dashed, (0.06,4.0) = gray solid. The eccentric case exhibits an oscillation with nodal spacings indicated by the dotted blue lines.}
\label{fig:wiggle_modes}
\end{figure}
In Fig.~\ref{fig:wiggle_modes} panel (a) we plot the absolute magnitude of the linear theory modal surface density perturbations $\delta\Sigma_{mm}$ for the case $(p,q)  = (1.5,0.0)$, normalised by the characteristic scaling $\delta\Sigma_{\rm lin}(r)$ given by equation (30) of \cite{CimermanEtAl_2024} as a function of $m$. In panel (b) we also plot the corresponding phase angle of this perturbation rotated by $\exp(im\phi_{\rm lin})$, where $\phi_{\rm lin}$ is the azimuthal position of the net density wake, tracking equations (10)-(14) of \cite{CimermanEtAl_2024}. In each panel, we compare two eccentricities $e = 0.0$ and $e = 0.06$, at two different radial locations, $r/a_{\rm p} = 3.0$ and $r/a_{\rm p}=4.0$. The corresponding line styles are indicated in the figure legend.

The upper panel is suggestive of a self-similar structure for the azimuthal wake profile since the normalised surface density perturbations at the two radii lie on top of each other for both eccentricities considered. However, the shape of the profiles clearly varies with $e$. For the circular case, $|\delta\Sigma_{mm}|$ decays monotonically, whilst the eccentric case exhibits an oscillation with nodal spacings indicated by the dotted blue lines. The self-similar behaviour is also borne out in the angular information shown in panel (b) where there is a coherent structure in the modal phases which smoothly evolve as the radius changes. Also note that the phases for both eccentricities perfectly overlap at the corresponding radii, modulo the $\pi$ discontinuity at the nodal lines signifying a sign flip as $\delta\Sigma_{mm}$ passes through zero.

The differing structures exhibited in panel (a) might initially lead one to surmise that the torque density response will be very different for the circular e=0.0 case versus the eccentric e=0.06 case. However, at large $r$ the dominant potential component is the low order $\Phi_{11}$ mode which is relatively insensitive to the eccentricity far away from the planet. This then couples to the $\delta\Sigma_{11}$ mode according to equation \eqref{eq:orbit_avg_amf_dtdr}. Fig.~\ref{fig:wiggle_modes} clearly shows that these low order modes are almost identical for the two eccentricities considered at large $r$. Therefore the torque wiggles are essentially preserved, even up towards moderate, transonic eccentricities. Indeed, whilst we have not plotted the corresponding $e=0.12$ curves in Fig.~\ref{fig:wiggle_modes} so as to maintain clarity, they essentially paint the same picture wherein there is close overlap with the $e=0.0$ curves for the lowest order modes.

However, the profiles observed for the larger eccentricities of $e = 0.3$ and $0.5$ begin to exhibit noticeable differences compared with the near circular runs. The high values of $e$, with larger planetary excursions, mean that off-diagonal and higher order modal contributions of the perturbing potential become ever more important in the far-field domain. Furthermore, the low order, diagonal density perturbations also begin deviating from the lower eccentricity curves plotted in Fig.~\ref{fig:wiggle_modes}. Together, this breaks the aforementioned reasoning and disturbs the wiggle structure. Presumably far enough away from the planet, the circular perturber wiggle profile can be reinstated, although the feature amplitudes will be very weak by this point. 

Finally, although the circular wiggle structure changes in this high $e$ regime, we continue to find good agreement between the simulations and linear framework. Although these methods do begin to present slight discrepancies near the planet, there appears to be effective overlap further away, underlining their success at capturing very subtle features. 

%---------------------------------%
\subsection{Integrated torques}
\label{subsec:integrated_torques}
%---------------------------------%

\begin{figure}
\includegraphics[width=\linewidth]{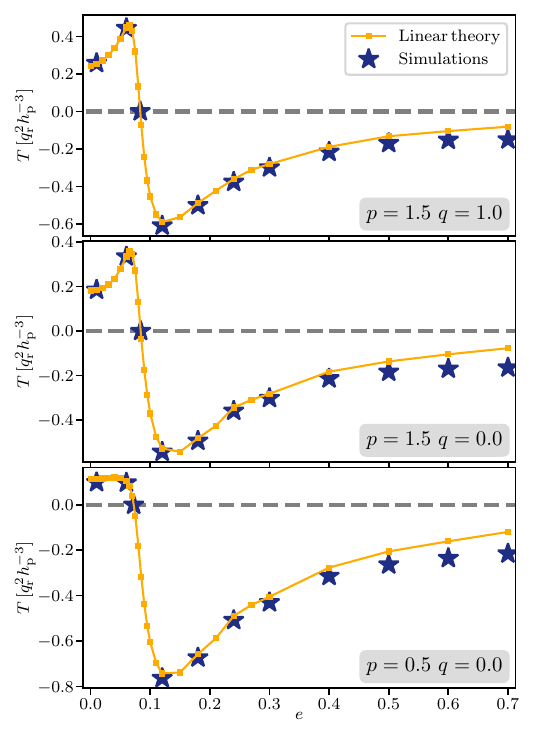}
\caption{Net integrated torque $T$ across the disc for the three disc profiles as a function of $e$. The linear theory results are shown as yellow markers with connective lines. The simulation results are denoted by blue starred markers.}
\label{fig:torque_pq}
\end{figure}

Whilst the torque density structure reveals information about the detailed radial coupling between the planet and the disc, it is the net integrated torque which ultimately governs the feedback onto the planet. To this end, we compute the net torque $T$ enacted across the disc for both our linear theory and simulations and plot the results in Fig.~\ref{fig:torque_pq}.
The three panels correspond to the different disc profiles being considered. Each shows the variation of the net torque as a function of $e$, extending from the circular regime to highly eccentric orbits with $e = 0.7$. The yellow markers and connective lines show the results from the linear theory calculations whilst the blue stars show the results from our suite of simulations. 

As anticipated by the agreement observed in the radial torque density curves examined in Figs.~\ref{fig:torque_p1.5q0.0} and \ref{fig:torque_p0.5q0.0}, here we also find close accordance in the integrated torque across all eccentricities and disc profiles. Notably, the simulations confirm the qualitative dependence on $e$ as previously discussed in detail by \FRII. Namely, for the steeper $p=1.5$ runs there is an initial enhancement in $T$ as one approaches and exceeds the transonic value of $e = h_{\rm p}$. The simulations are specifically targeted to capture this peak torque around $e = 0.06$, finding almost perfect overlap with the linear predictions. Meanwhile we see that for the shallower surface density profile $p=0.5$, this enhancement is less pronounced and the $T$ remains relatively flat up towards $e = h_{\rm p}$. Beyond this transonic point, we observe a clear transition in behaviour as the net torque undergoes a sharp reversal becoming highly negative and minimised when $e\sim 2h_{\rm p}$. This corresponds to the value of eccentricity for which the planet moves supersonically with respect to the gas at all orbital phases. 

It was at this value of $e=0.12$ that \FRII truncated their study due to uncertainty in the modal convergence for higher eccentricities. However, here we push much further and extend our simulations and linear theory up towards $e=0.7$. The net torque remains negative but slowly decreases in absolute value as $e$ increases. Fitting a power law between $e = 0.18$ and $e=0.7$ to our linear theory results, yields the best fitting exponents $(-1.1,-1.1,-1.0)$ for the disc profiles $(p,q) =$ $(1.5,1.0)$, $(1.5,0.0)$ and $(0.5,0.0)$ respectively. This scaling is close to the heuristically motivated fitting functions presented in \cite{IdaEtAl_2020}, which suggest that $T \propto e^{-1}$ in the highly supersonic limit. Once again, the broad trends are in good agreement between the methodologies. However, for the highest values of $e$ we begin to see larger discrepancies emerge as the linear theory underestimates the torque magnitude found in the simulations. To test this further, we have performed a convergence study in appendix \ref{app:convergence} which demonstrates that our linear framework and simulation torque diagnostics do not change significantly upon including more modes or doubling the fiducial resolution. This suggests that the differences arise due to more intrinsic difficulties. Perhaps the high value of $e$ fundamentally precludes convergence of our modal scheme. Furthermore, the very oscillatory nature of the wave contributions for high order modes becomes mutually challenging for the linear theory ODE solvers and the fully numerical hydro scheme. Whilst the large radial planetary excursions potentially lead to more troublesome interaction with the boundaries, we show in appendix \ref{subsec:convergence_remarks} that the differences are primarily sourced between apocentre and pericentre. The intricate torque structures in this interval might point towards subtle nonlinear effects that affect the disc response in this region. For example, in section \ref{subsec:amf_deposit} we show that nonlinear wave steepening can deposit angular momentum into the disc, which will modify the background density structure slightly, potentially altering the interaction with the planet.

%---------------------------------%
\subsection{Orbital evolution}
\label{subsec:orbital_evolution}
%---------------------------------%

The back-reaction of the net disc torque yields an equal and opposite torque on the planet, acting to modify its angular momentum. Furthermore, the power enacted on the planet governs the change in its orbital energy. Together, this combination of torque and power govern the evolution of orbital elements in accordance with equations \eqref{eq:timescale_relation} and \eqref{eq:inverse_timescales}. Notably, torque reversals alone do not suffice to change the direction of orbital migration, which depends on the power sapped from or delivered to the planet, while the interplay between torque and power determine the evolution of planet's eccentricity.

\begin{figure}
\includegraphics[width=\linewidth]{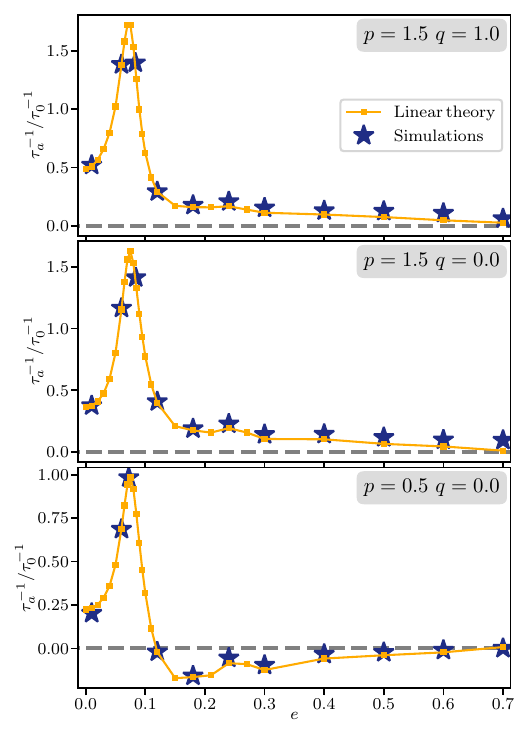}
\caption{Characteristic semi-major axis evolution rates $\tau_{a}^{-1}$, as a function of $e$, for the three disc profiles. Linear theory (yellow markers with connective lines) versus simulation results (blue stars). }
\label{fig:taua}
\end{figure}

\begin{figure}
\includegraphics[width=\linewidth]{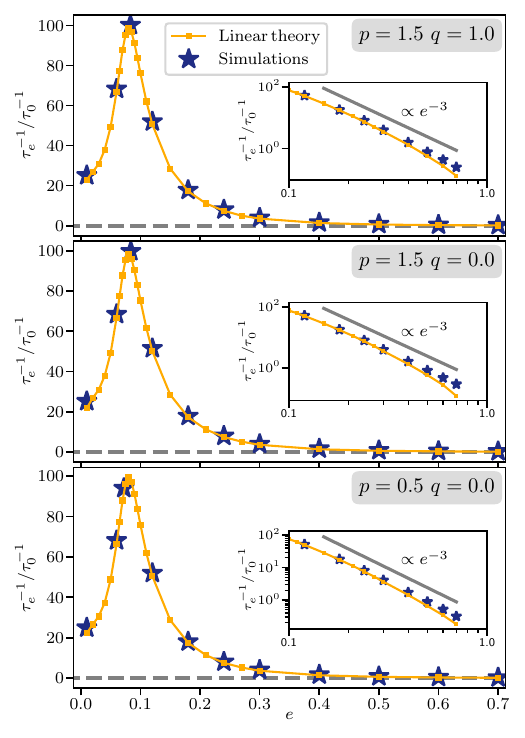}
\caption{Same as Fig.~\ref{fig:taua} but for $\tau_e^{-1}$. We are also excluding the point where $e$ is identically equal to $0.0$ for which the characteristic eccentricity damping rate is ill-defined. Meanwhile, the inset panels show the curves at high $e$ on logarithmic axes.}
\label{fig:taue}
\end{figure}

In Figs.~\ref{fig:taua} and \ref{fig:taue} we examine this orbital element interplay for $\tau_a^{-1}$ and $\tau_e^{-1}$ respectively, as a function of $e$. Once again, the three rows in each figure correspond to the different disc profiles. Meanwhile the yellow markers and connective lines correspond to the linear theory predictions, whilst the blue stars mark the simulation results. The results are plotted in units of the characteristic migration rate $\tau_0^{-1} \equiv F_{J,0}/(M_{\rm p}n_{\rm p} a_{\rm p}^2)$.


Focusing first on the semi-major axis evolution in Fig.~\ref{fig:taua}, we again note the very good agreement between simulations and linear theory across a wide range of $e$. In particular we observe the initial enhancement in the $\tau_a^{-1}$ damping rate as $e$ crosses the transonic value. This actually peaks at slightly larger $e \sim 0.07-0.08$, compared with the net torque curves examined previously. Subsequently, $\tau_a^{-1}$ turns over and is strongly suppressed as the orbit becomes entirely supersonic by $e = 0.12$. Even though this is associated with a net torque reversal, we see that for the disc profiles with $p=1.5$, $\tau_a^{-1}$ remains positive. However, for the density profile with $p=0.5$, we see that the inwards migration does in fact initially stall at $e=0.12$ before reversing above this. The simulation results closely trace this behaviour and verify the findings of \FRII that supersonic eccentric planets in shallow surface density discs are susceptible to outwards migration. Increasing $e$ towards the upper value of $0.7$ shows the magnitude of $\tau_a^{-1}$ tends towards 0 and the migration rate, inwards or outwards, is reduced. Interestingly, this trend is non-monotonic and one can discern subtle bumps in the curves about $e \sim 0.27$. This feature is found in both our simulations and the linear theory, lending weight to their validity. At the highest values of $e$ considered we begin to note discrepancies between the yellow markers and stars. Although the migration rates are both small here, the relative differences are becoming more significant. It is at this point we curtail our push towards higher $e$, since the simulations and linear theory both encounter challenges which affect their reliability (see section \ref{subsubsec:higher_eccentricities}).

Now turning our attention to Fig.~\ref{fig:taue}, we examine the behaviour of the characteristic eccentricity damping rate $\tau_e^{-1}$. Firstly, note the different y-axis scale which is roughly two orders of magnitudes larger compared with Fig.~\ref{fig:taua}, consistent with our expectation that the $\tau_e^{-1}$ is much faster than $\tau_a^{-1}$ \citep[as found previously by][]{PapaloizouLarwood_2000, TanakaWard_2004}. Once again we see the simulations and linear results are in very good agreement. The characteristic eccentricity damping exhibits a sharp rise which peaks around $e\sim 0.08$. Beyond this enhancement we see a steep suppression which is consistent with $\tau_{e}^{-1}\propto e^{-3}$, as predicted by \cite{PapaloizouLarwood_2000}. As $e$ tends towards $0.7$ the damping rate converges towards zero. In order to better interpret the variation in these high $e$ cases, the  subplots also show the variation of $\tau_e^{-1}$ on a logarithmic scale. This better exhibits the power law behaviour, with the predicted $\propto e^{-3}$ scaling indicated by the inset grey lines. Further connections between our present results and previous predictions will be discussed in detail in section \ref{subsec:connection}

%---------------------------------%
\subsection{AMF and torque deposition function}
\label{subsec:amf_deposit}
%---------------------------------%

Up to this point we have focused our attention on the torques which communicate angular momentum between the planet and the disc, causing orbital evolution. However, it should be noted that the angular momentum exchanged with the gas is not immediately deposited into the background disc. Instead the coupling is mediated by the launching of density waves, as visualized in Fig.~\ref{fig:sigma_grid_exps}, which propagate away from the planet and transport angular momentum beyond the excitation region. It is only when the waves damp via linear \citep[][]{Takeuchi1996,MirandaRafikov_2020} or nonlinear \citep[][]{GoodmanRafikov_2001,Rafikov2002a,CimermanRafikov_2021} processes that they deposit angular momentum causing the disc to evolve -- driving mass flow and sculpting structures such as gaps and bumps in the surface density profile \citep[e.g.][]{Lynden-BellPringle_1974,Rafikov_2002, DempseyEtAl_2020, CordwellRafikov_2024}. 

Our linear theory is manifestly only able to capture the linear evolution of excited density waves. For our (locally) isothermal discs a freely propagating density wave is associated with a conserved wave-action \citep[e.g.][]{LeeLiYongJie_2016, MirandaRafikov_2020} such that
\begin{equation}
    \label{eq:conserved_AMF}
    \frac{d}{dr}\left(\frac{F_J}{c_s^2}\right) = 0.
\end{equation}
In a globally isothermal setting, this means that the AMF is conserved far from the planet and no angular momentum is exchanged with the background disc. Meanwhile, for $q>0$ the AMF decays in the outer disc and rises in the inner disc as the instantaneous cooling leads to additional source/sink terms in the wave energy equations \citep[as elucidated by][]{OgilvieEtAl_2025}. This can deposit or extract angular momentum from the disc, leading to an `anomalous' mass flux, as described by \cite{MirandaRafikov_2020}. 

To simplify our analysis in this section, we will focus on globally isothermal runs for which the linear theory predicts AMF conservation. In this thermal regime, the change of $F_J$ beyond the excitation region can only be facilitated by nonlinear effects. In particular, \cite{GoodmanRafikov_2001} and \cite{Rafikov2002a} showed that the waves excited by a circularly orbiting perturber steepen as they propagate until shocking at a distance
\begin{equation}
    l_{\rm sh} \approx 0.8 H_{\rm p}\left(\frac{\gamma+1}{12/5} \frac{M_{\rm p}}{M_{\rm th}}\right)^{-2/5},
\end{equation}
away from the planet, at which point they start to deposit their angular momentum. Here, $\gamma$ is the adiabatic index which is set to be 1 in the context of our locally isothermal study. This shocking behaviour is naturally captured by fully nonlinear, numerical simulations. However, one should note that simulations are also susceptible to numerical dissipation which can artificially damp the waves. At our fiducial lower resolution, damping associated with weakly nonlinear shocks was precluded by numerical dissipation , although the eccentric modification of $F_{J}$ associated with planetary excursions and the mutual interaction of waves near $r_{\rm p}$ was resolved adequately. Therefore, in order to reduce numerical dissipation, in this section we repeated a limited set of high resolution simulations with much higher radial resolution $N_r=18432,$ but holding the azimuthal resolution constant at $N_\phi\approx6528$. In the vicinity of the planetary semi-major axis, this corresponds to $\sim 281$ cells/$H_{\rm p}$.\footnote{This resolution approaches that of \cite{CimermanRafikov_2021}, which used 464 cells/$H_{\rm p}$ in order to achieve quantitative convergence of the shock formation. However, their planet mass was $M_{\rm p} = 0.01 M_{\rm th}$ and they only considered circular orbits, precluding a direct comparison.}

\begin{figure*}
    \centering
    \includegraphics[width=\linewidth]{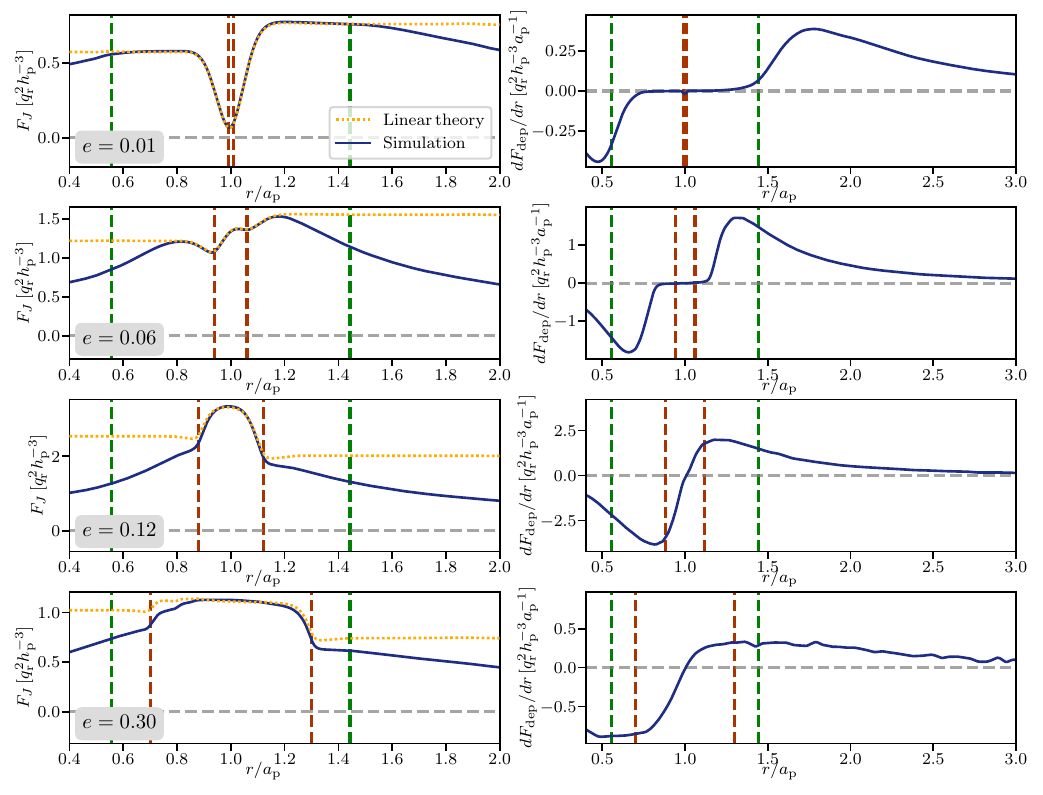}
    \caption{Nonlinear deposition of angular momentum. \textit{Left column}: radial profiles of AMF for linear theory (yellow dotted) and simulations (blue solid). \textit{Right column}: the radial profile of the deposition torque (defined by equation \eqref{eq:deposition_trq}) measured from the simulations. The dashed vertical lines denote the planetary apo/pericentres (brown) and the characteristic circular planet shocking locations $a_{\rm p} \pm l_{\rm sh}$ (green). Moving down the rows we plot the results for different eccentricities $e \in \lbrace 0.01,0.06,0.12,0.3\rbrace$}.
    \label{fig:amf_deposit}
\end{figure*}

To this end, we visualise the AMF variation across the disc in Fig.~\ref{fig:amf_deposit}. We focus on the simplest globally isothermal case with $q=0.0$ so that, within linear theory, $F_J$ is conserved. Furthermore, we adopt $p=1.5$ so there are no AMF jumps associated with vortensity gradients which drive corotation torques. In the left hand column we plot $F_J$ as a function of radius for both the linear theory (yellow dotted) and simulation (blue solid) results. Moving down the rows we consider four values of $e \in \{0.01,0.06,0.12,0.30\}$. 

In the near circular case, we see the classic picture wherein the AMF rises steeply about the planetary orbit as waves are launched predominantly from Lindblad resonances within the vicinity of $|r-a_{\rm p}| \sim H_{\rm p}$. As required by AMF conservation, the curves corresponding to linear theory plateau further away from the planet where the coupling to the potential is weak and the waves propagate freely. However, the simulation clearly deviates from this as $F_J$ begins to decay due to nonlinear shocking. Encouragingly, we see that the onset of this departure coincides with the predicted shocking locations $a_{\rm p}\pm l_{\rm sh}$, denoted by the dashed green lines. This suggests that the resolution is high enough to suppress the numerical dissipation which causes the AMF to start decaying closer to the planet in our fiducial resolution runs.

As the eccentricity is increased, the radial structure of the excited AMF notably changes. For the transonic value of $e=0.06$ we see this as a series of small `steps' between the peri and apocentre locations, denoted by the brown dashed lines. Within this interval, the linear theory and simulations still agree fairly well. Once again, further out the simulation AMF decays whilst the linear theory predicts that $F_J$ is preserved. However, now it appears that the nonlinear damping sets in closer to the planetary semi-major axis. As we consider the fully supersonic regime with $e = 0.12$, we see that the AMF about $a_{\rm p}$ actually flips to become a local maximum. This enhancement might be interpreted as the superposition of the inwards travelling waves launched from apocentre and the outwards travelling waves excited at pericentre. Beyond this, the linear theory now saturates with a higher level of $F_J$ in the inner disc compared with the outer disc. This inversion is consistent with the torque reversals discussed in section \ref{subsec:integrated_torques}. Meanwhile the simulations again exhibit a clear departure from the linear theory outside of the orbital extremities. The picture is similar for $e=0.3$ where the central peak is now more spread out due to the larger radial excursions executed by the planet. Since the excitation happens over a wider range, there is also more scope for waves to shock and dissipate before propagating out of this epicylic interval. Therefore the AMF actually begins to dip beneath the linear theory predictions within the radial range of the planetary excursions as one examines close to pericentre and apocentre.

Another useful metric to quantify the nonlinear deposition of AMF is visualized in the right hand column of Fig.~\ref{fig:amf_deposit}. Here we plot the deposition torque density
\begin{equation}
    \label{eq:deposition_trq}
    \frac{d F_{\rm dep}}{dr} = \frac{d T}{dr}-\frac{dF_J}{dr},
\end{equation}
which accounts for the balance between the torque density injected by the perturber and the radial variation of wave AMF.\footnote{In these plots we restrict out attention to $r/a_{\rm p} > 0.4$. Interior to this the interaction with the inner boundary contaminates the AMF and the behaviour is less reliable.} In purely linear theory these terms cancel and there is no angular momentum deposited in the background disc. With nonlinear shocking at play, $-dF_J/dr$ dominates the deposition. Knowledge of the spatial distribution of $dF_{\rm dep}/{dr}$ is important for modelling the evolution of surface density, as expounded in the framework developed by \cite{CordwellRafikov_2024}. 

In the upper panel, for $e=0.01$ we see the expected shape for this profile, broadly consistent with previous circular studies (e.g. Fig.~4 of \cite{CimermanRafikov_2021}). Namely, near the coorbital location the curve is flat before increasing about the predicted shocking location $a_{\rm p} \pm l_{\rm sh}$. As the eccentricity increases to 0.06, the coorbital flat portion shrinks, the rise of $dF_{\rm dep}/dr$ begins just outside the peri/apocentre locations and the deposition peaks move interior to the circular planet shocking location. For the transonic values of $e = 0.06$ and $0.12$ there is a strong enhancement in the magnitude of the deposition function. This is in part fuelled by the spike in torque densities (see Fig.~\ref{fig:torque_p1.5q0.0}) which excite larger wave AMFs which can then be deposited. Furthermore, when the planet moves at speeds $\sim c_s$, the excited wave can pile up near the chasing planet -- this increased response naturally leads to stronger shocks. Interestingly, for the supersonic values of $e$ the deposition peaks actually coincide with the pericentre and apocentre locations. Between these locations $dF_{\rm dep}/{dr}$ follows an approximately linear slope. For $e = 0.3$ we see that as the planetary epicycle widens the peaks start moving further out again. Furthermore, we see a return to smaller deposition amplitudes as the excited AMF is reduced and the shock strength weakens.
% ================================%
\section{Discussion}
\label{sec:discussion}
% ================================%

% -------------------------------%
\subsection{Connection with previous work}
\label{subsec:connection}
% -------------------------------%

The excellent agreement between the linear theory and our numerical simulations validates the results of \FRII, which explored up towards moderately supersonic values of $e$. In this prior work, their linear methodology (same as presented here) was suggested to fit the previously available numerical simulations \citep[e.g.][]{CresswellNelson_2006} much better compared with the convenient but somewhat heuristic fitting prescriptions stitched together across the literature \citep[e.g.][]{PapaloizouLarwood_2000,TanakaWard_2004,MutoEtAl_2011,IdaEtAl_2020}. Indeed, Fig.~12 in \FRII points towards the shortcomings of these previous fits in simultaneously capturing the position of the torque reversal, the enhancement in orbital element damping rates around the transonic value of $e$ and accurately reproducing the limiting behaviour in both the low and high eccentricity regimes. Meanwhile, our tailored simulations nicely recover all of these predictions of linear theory whilst offering a more rigorous benchmark for future studies. Accordingly, in appendix \ref{app:code_test} we have provided a repository of our linear torque data which can be used as a robust test for new planet-disc code setups.

Our current study also pushes towards high $e$ values where we continue to find good correspondence between our simulations and linear theory. In the past this supersonic regime has received some attention with varying levels of agreement compared with our new results. \cite{PapaloizouLarwood_2000} (hereafter PL00) explored the eccentric planet-disc problem by expanding the perturbing potential, akin to our modal approach. However, they evaluate the torque contribution from each resonance by a purely local WKB formula, akin to \cite{GoldreichTremaine_1979}. Furthermore they only explore a single disc structure with $(p,q) = (1.5,1.0)$ which makes it difficult to generalize their results. 

Alternative approaches have often invoked a dynamical friction formalism to calculate the gravitational drag acting on the planet. For example \cite{ONeillEtAl_2024} have computed the wake structure sourced by an eccentric perturber in a homogenous and isothermal medium. However, drawing detailed comparisons with the planet-disc problem is unclear since the differential rotation of the background shear flow is not included.  A more compatible study was undertaken by \cite{MutoEtAl_2011} (hereafter MTI11) which considered the relative velocity between the eccentric perturber and a background Keplerian disc. They inserted this relative velocity into a drag formalism applicable for highly supersonic motion \citep[inspired by][]{Ostriker_1999}, to evaluate the force on the planet around the orbit. The resulting orbit averaged evolution of orbital elements was explored across disc parameters and fitting functions are supplied. It should be cautioned that such a method assumes the steady state dynamical friction response is established instantaneously at each point around the orbit, whilst ignoring the wavelike response of the disc which is excited in a resonant fashion.

\begin{figure}
    \centering
    \includegraphics[width=\linewidth]{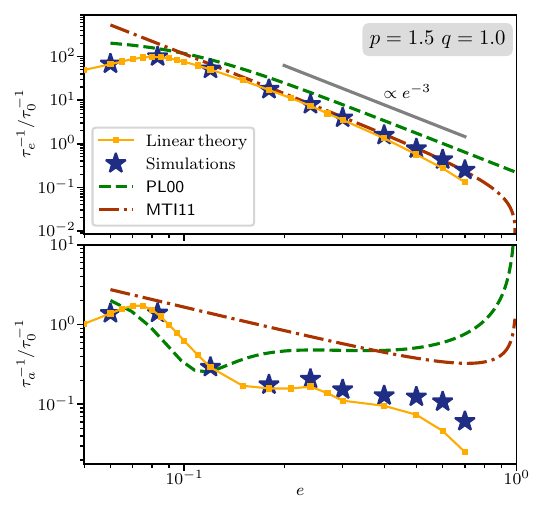}
    \caption{Comparing orbital evolution prescriptions with our results for a disc with $(p,q) = (1.5,1.0)$ as a function of $e$. \textit{Upper panel}: $\tau_e^{-1}$ rate. \textit{Lower panel}: $\tau_a^{-1}$ rate. The lines correspond to the linear theory (yellow with square markers), simulations (blue stars), PL00 prescription (green dashed) and the MTI11 fitting function (magenta dashed).}
    \label{fig:theory_comparison}
\end{figure}

In order to benchmark these previous methods, we compare the available fitting functions to our new results in Fig.~\ref{fig:theory_comparison}. The upper and lower panels show $\tau_e^{-1}$ and $\tau_a^{-1}$ respectively, as functions of $e$, for the disc profile $(p,q) = (1.5,1.0)$. Note the logarithmic scale spacing is chosen to emphasise any power-law behaviour (or lack thereof) in the high $e$ regime. As per usual, our linear theory and simulations results are denoted by the yellow lines/markers and the blue stars respectively. The green dashed lines correspond to the fitting functions proposed by PL00, whilst the brown dash-dotted lines show the predictions of MTI11.
We see that both eccentricity damping prescriptions agree reasonably well with our results. In particular the dynamical friction predictions of MTI11 closely overlap the simulation results up to the limiting values of $e$ considered. This is also true for the linear theory up until $e \gtrsim 0.4$, at which point there is a slight steepening of the yellow markers. Nonetheless, the trends are broadly consistent with the $\tau_e^{-1}\propto e^{-3}$ relationship, originally predicted by PL00, as indicated by the lines lying parallel to the reference grey slope bearing this power law profile. Contrastingly, in the lower panel we see that the fitting functions provide much poorer fits for $\tau_a^{-1}$. They fail to capture the non-monotonic structure around $e = 0.24$, the general trend towards high $e$, and the magnitude of the migration rate. This suggests that whilst these simpler prescriptions might be appropriate for the eccentricity damping rate, one should be more cautious when invoking the corresponding migration rates.

% -------------------------------%
\subsection{Caveats}
\label{subsec:caveats}
% -------------------------------%

Despite the success of our targeted study, there are several caveats that present opportunity for further investigation. Indeed, we have adopted a number of simplifications which we will briefly address here. Namely we will consider nonlinear corotation effects, higher planetary masses, different disc parameters and even higher eccentricities.

% ...............................%
\subsubsection{Nonlinear corotation effects}
\label{subsubsec:corotation}
% ...............................%

It should be noted that general disc profiles set up a non-zero vortensity gradient and are therefore susceptible to corotation torques \citep[][]{GoldreichTremaine_1979}. Since the disc response about corotation is evanescent, the angular momentum deposited in the vicinity of these locations is unable to propagate away. Indeed, the exchange of angular momentum with the disc here can change the background profile by setting up horseshoe trajectories \citep[][]{Ward_1991}. This fundamentally nonlinear effect is not captured by linear theory \citep[][]{OgilvieLubow_2003,BrownOgilvie_2024a} and might lead to discrepancies. However, as investigated by \cite{PaardekooperPapaloizou_2009a}, the unsaturated horseshoe region takes some fraction of the libration timescale to form, before saturating on a longer libration timescale. Meanwhile the linear corotation torque transiently grows to its steady state over shorter dynamical timescales. For our small, simulated planet-to-star mass ratio $q_{\rm r} = 10^{-6}$, the order of magnitude horseshoe width is given by 
\begin{equation}
    \frac{x_{\rm s}}{a_{\rm p}} \sim \sqrt{\frac{2q_{\rm r}}{3 \epsilon}} = 6\times10^{-3},
\end{equation}
as suggested by \cite{PaardekooperPapaloizou_2009a}. This corresponds to a characteristic libration timescale
\begin{equation}
    \tau_{\rm lib} \sim \frac{8\pi}{3 x_{\rm s}}n_{\rm p}^{-1} \simeq  220 T_{\rm p}.
\end{equation}
Therefore, up until now, by sampling our simulation torques at times $t$ satisfying $T_{\rm p} < t \ll \tau_{\rm lib}$ we have aimed to probe the torques during a phase where they have had time to establish the linear response, but before saturation effects complicate proceedings. However, we now run a selection of extended runs up towards 200 planetary orbits in order to start probing these late time, nonlinear effects.

\begin{figure}
    \centering
    \includegraphics[width=\linewidth]{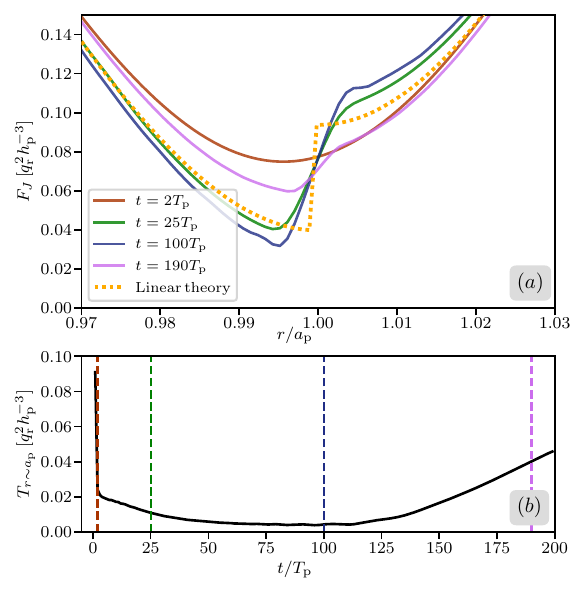}
    \caption{AMF and torque evolution in the coorbital region for $e=0.01$. \textit{Panel (a)}: shows the AMF radial profile for the simulation results at different times denoted by the solid, coloured lines. The linear theory prediction is shown as the overlying dotted yellow line. \textit{Panel (b)}: the coorbital torque $T_{r\sim a_{\rm p}}$ contribution is plotted as a function of time. The coloured, vertical dashed lines correspond to the time slices shown in (a). Note that the y-axes here are much narrower than the typical AMF and torque scales used in previous figures, in order to highlight these subtle coorbital effects.}
    \label{fig:corotation_e0.01}
\end{figure}
\begin{figure}
    \centering
    \includegraphics[width=\linewidth]{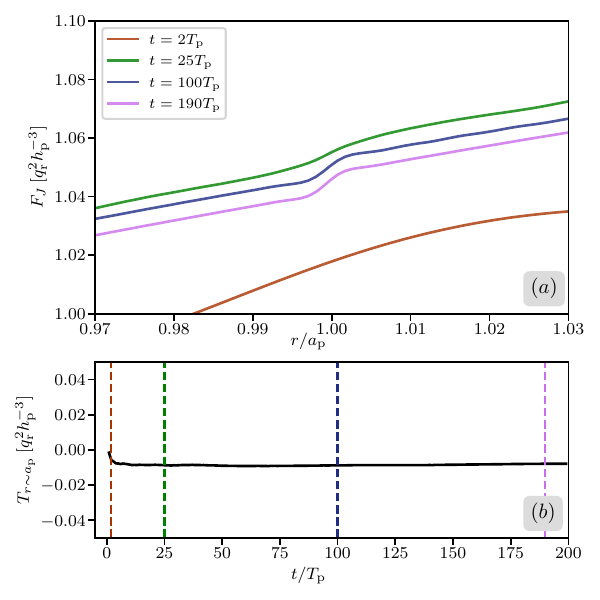}
    \caption{Same as Fig.~\ref{fig:corotation_e0.01} but for $e = 0.30$. Also note that the linear theory curve lies beyond the plotted range.}
    \label{fig:corotation_e0.3}
\end{figure}

In Fig.~\ref{fig:corotation_e0.01} we first visualise this time dependence for the simplest case of our near circular run with $e = 0.01$, in the disc with $(p,q) = (0.5,0.0)$. In panel (a) we zoom in to the AMF distribution around the coorbital region between $|r-a_{\rm p}|<0.03$. The solid lines show the $F_J$ profiles measured in our simulation at different times. The colours correspond to the time slices shown in panel (b), which in turn show the time dependence of the disc integrated torque $T_{r\sim a_{\rm p}}$ within this narrow radial range\footnote{Note that this integral will pick up contributions from both Lindblad and corotation resonances.}. We clearly see that $T_{r\sim a_{\rm p}}$ undergoes a long time scale evolution. At $t=T_{\rm p}$ the AMF profile is rapidly evolving and the net torque profile is still settling down. About the coorbital region, $F_J$ is smooth at this time. However, over several dynamical timescales the AMF develops a step-like jump which is comparable to the predictions from linear theory (see the over-plotted dotted line). Within the context of linear theory, this flux `discontinuity' is a measure of the corotation torque \citep[][]{GoldreichTremaine_1979,KorycanskyPollack_1993} and has previously been used to partition the net torque into its Lindblad and corotation contributions (e.g. \FRII). However, we should note that the torque and the AMF jump profile continue to slowly evolve, with $T_{r\sim a_{\rm p}}$ decreasing towards a minimum by $t \sim 75 T_{\rm p}$. This represents the slow transition towards the unsaturated horseshoe drag regime, over a fraction of $\tau_{\rm lib}$, where there is an enhancement in the negative corotation torques, qualitatively similar to the picture presented in Fig.~7 of \cite{PaardekooperPapaloizou_2009a}. At $t=100T_{\rm p}$, during this unsaturated phase, the AMF jump looks similar to the earlier times but with a slightly larger magnitude, consistent with the decrease in disc torques. Beyond this time, libration effects begin to enter as material has had time to traverse the tadpole orbits and catch up with the planet. We see that the net torque begins to rise as saturation starts setting in. Simulating for longer times would presumably show the long timescale libration oscillations similar to those seen in Fig.~12 of \cite{PaardekooperPapaloizou_2009a}, before levelling off in the saturated state. We see that this phase of evolution is associated with a flattening out of the AMF and the jump is suppressed.

We compare this behaviour with a run boasting the much higher eccentricity of $e=0.3$. The corresponding results are shown in Fig.~\ref{fig:corotation_e0.3}. Once again there is an initial transient phase where the torques and AMF profiles are rapidly set up over dynamical timescales. However, by $t=25T_{\rm p}$ the $F_J$ discontinuity has formed but with a much smaller amplitude. Furthermore, this profile seems to undergo slower evolution compared with the near circular run discussed above. We find that the AMF curves closely overlap for all $t \geq 25T_{\rm p}$ and the $T_{r\sim a_{\rm p}}$ profile remains level towards late times. This modified picture is consistent with the view that larger eccentricities suppress the corotation torques. This has previously been seen within the context of linear theory by \FRII and also found for the nonlinear horseshoe drag regime by \cite{FendykeNelson_2014}. They attribute this effect to the enhanced softening scale set by the larger epicylic excursions of the planet, which in turn shrinks $x_{\rm s}$. This will increase $\tau_{\rm lib}$ and hence explains the delayed onset of saturation signatures in our more eccentric runs. Within the context of linear theory this might be interpreted as the dilution of the perturbing potential over many more, non-coorbital corotation resonances, which weakens the primary coorbital resonance. These non-coorbital corotation resonances are also predicted to exhibit small AMF jumps within the context of the linear theory (see \FRII, Fig.~7) although these very subtle features are not found in our simulations. Indeed, the extended effects of nonlinear shocking lead to stronger departures from the linear theory in the region between peri and apocentre, as discussed in section \ref{subsec:amf_deposit}. This complicates the comparison with linear AMF predictions and presumably washes out many of these subtle features.

% ...............................%
\subsubsection{Higher mass planets}
\label{subsubsec:higher_mass}
% ...............................%

In this study we have focused our attention on low mass planets so as to remain confidently within the linear regime. Our choice of $h_{\rm p} = 0.06$ and the mass ratio $q_{\rm r} = 10^{-6}$ gives a planet mass of $M_{\rm p} = 0.0046 M_{\rm th}$. For a solar mass star, this translates into $M_{\rm p} = 0.33M_{\oplus}$ and is therefore most relevant to terrestrial mass planets or stellar mass black holes in AGN discs. However, for larger planetary masses we expect the picture to change. Indeed, during our investigation we also tested runs with $q_{\rm r} = 10^{-5}$ (not shown). This mass ratio is still firmly sub-thermal and the main torque profiles remain the same (although amplitudes are scaled by a factor of 10 compared with our fiducial runs). However, we do find that there are more noticeable discrepancies between the torque wiggles predicted by linear theory and simulations. Clearly, these subtle features will be modified by even small amounts of nonlinearity.

Increasing the planetary mass further would result in more pronounced differences which we do not consider in this paper. \cite{ZhuZhang_2022} performed exemplary 2D simulations of eccentric Jupiter mass planets, showing that the spiral wake morphology is significantly modified compared with the low mass regime. Namely, the larger amplitude wakes can interact nonlinearly, leading to the messier interference of wake crossings. Furthermore, the rapid steepening and shocking of the waves deposits angular momentum more efficiently. This can promote gap opening and a transition towards Type II migration. Interestingly, as the eccentricity increases, the gap depth is found to be shallower \citep[also see][]{DuffellChiang_2015} which might help constrain the properties of hypothesized planets in observed gaps. 

The feedback on the planetary orbital elements in these gapped discs can also lead to significant changes in orbital behaviour. The depletion of material in the gap can suppress coorbital resonances which tend to damp eccentricity \citep[][]{Goldreich2003}. Simulations by \cite{DuffellChiang_2015} have shown that the eccentricity can then be excited and sustained around $e \sim h_{\rm p}$. Contrastingly, we have found that in the linear regime, this transonic eccentricity is actually associated with enhanced damping. Subsequent high-mass simulations by \cite{Ragusa2018} even suggest that $e$ can reach even larger values as the disc and planet exchange eccentricity over secular timescales. Understanding this richness of behaviour continues to be an avenue for exploration and in future work it would be instructive to systematically explore the variation in torque profiles and as a function of planet mass and eccentricity.

% ...............................%
\subsubsection{Different disc parameters}
\label{subsubsec:disc_parameters}
% ...............................%

In this study we have limited our attention to an an illustrative set of disc profiles to demonstrate the correspondence between between linear theory and simulation across a representative cross-section of parameter space. Indeed, our three chosen combinations of $p$ and $q$ probe discs both with and without temperature and vortensity gradients. In \FRII a finer grid of $(p,q)$ parameter space was explored using the linear theory, but only up to a maximal eccentricity of $e = 0.12$.

It should also be acknowledged that here we have focused our attention on fiducial values for the scale height and the softening parameter. In reality, the results will be sensitive to both of these parameters too. Indeed, we have seen that the wake response exhibits a clear transition in behaviour when the planet is subsonic versus supersonic -- as determined by the characteristic ratio $e/h_{\rm p}$. This was demonstrated in \FRII wherein the $e$ value of the torque reversal scaled with the aspect ratio. Meanwhile, the heuristic softening prescription presents a more fundamental ambiguity for the 2D formalism. \cite{PaardekooperEtAl_2010} point out that the torque results can vary strongly as a function of the softening parameter, whilst \cite{DAngeloLubow_2010} warn that such smoothing cannot simultaneously capture all aspects of the intrinsic, un-softened 3D torques. A more rigorous vertical averaging procedure with a self-consistent softening prescription has recently been developed by \cite{BrownOgilvie_2024a} which could help remove some of this ambiguity. Furthermore, upgrading the linear formalism towards three-dimensions, akin to the approach developed by \cite{TanakaOkada_2024}, would be a powerful extension to our methodology.

Ultimately, this complex interplay between $(p,q,h_{\rm p}, \epsilon,e)$ in setting the net torque precludes a simple fitting function across a wide range of parameter space. Accordingly \FRII instead shared their results in a data cube for more flexible interpolation.

% ...............................%
\subsubsection{Higher eccentricities}
\label{subsubsec:higher_eccentricities}
% ...............................%

Whilst we push to highly supersonic values of $e$, we curtail our study at the upper value of $e=0.7$. By this point we are beginning to find small but noticeable discrepancies in the linear theory and simulation results. For example, the amplitude of sharp radial torque density features are not in exact agreement in Figs.~\ref{fig:torque_p1.5q0.0}-\ref{fig:torque_p0.5q0.0} for $e = 0.5$, whilst the linear theory seems to slightly underestimate the magnitude of numerical net torques in Fig.~\ref{fig:torque_pq}. However, in appendix \ref{subsec:linear_convergence} we have demonstrated sufficient modal convergence in $|m-l|$ for the fiducial upper value of $m = 170$ adopted by the linear framework. We also tested the use of expanded ranges with $m_{\rm max}=220$ and $|m-l|_{\rm max}=80$, but this did not provide a significant improvement on the comparison with simulations. 

Additionally, in appendix \ref{subsec:simulation_convergence} we also tested the convergence of our simulations with increasing resolution. We compared our fiducial torque diagnostics for our $(p,q) = (0.5,0.0)$ and $e=0.5$ run against a double resolution simulation, finding no substantial differences. This suggests that the discrepancy between simulation and theory owes to more fundamental differences.

One might surmise that both methodologies are complicated by stronger interactions with the boundaries as the radial planetary excursions become very large. However, by examining where the torque errors accrue in appendix \ref{subsec:convergence_remarks} we find that the discrepancy is predominantly sourced between the planetary peri/apocentre where the torque profile is particularly complex. This might point towards the need for an infeasibly high number of modes with a very slow convergence rate. Alternatively, there may be intrinsically nonlinear effects captured by the simulations which are causing the differences. For example, we have seen in section \ref{subsec:amf_deposit} that nonlinear steepening leads to shock deposition of angular momentum. For large eccentricities the planetary position can overlap the regions where AMF is deposited. This might change the background structure of the disc, altering the planet-disc coupling and the torque interaction. These possibilities will be explored in future work.

\if 0
Furthermore, upon testing the linear theory for higher values of $e>0.7$, the $m=1$ modes often give spurious results hinting at the breakdown of this approach. The erroneous behaviour might be related to the fact that inner Lindblad resonances formally move to $r = 0$ for $m=1$, and hence radiative inner boundary condition can misbehave (as discussed in \FRII). Exploring ways to stabilise the scheme in these cases might be pursued to try bring the simulations and linear theory into better alignment. 
\fi
% ================================%
\section{Conclusions}
\label{sec:conclusions}
% ================================%

In this paper we have performed a targeted set of hydrodynamical simulations to test the global, linear, 2D theory of eccentric planet-disc interactions, previously developed by \FRI and \FRII. We have placed a particular emphasis on connecting the low and high eccentricity regimes, spanning a range of $e$ values between 0.0 and 0.7, for which the planet moves subsonically and significantly supersonically with respect to the background Keplerian flow. We adopt simple, locally isothermal power law disc profiles and investigate a representative set of density and temperature structures. We examine a range of comparative diagnostics between the linear theory and simulations including wake morphologies, radial torque density profiles, net integrated torques, orbital evolution rates and angular momentum fluxes. Together, these results demonstrate that both approaches are in fantastic agreement, whilst also highlighting a number of key physical takeaways.

Indeed, not only are the qualitative wake structures in good correspondence, but the detailed radial torque density profiles also overlap remarkably well -- even the subtle torque wiggles are captured in the inner and outer disc. As $e$ becomes large, the linear theory and simulations both resolve the fine structure excited between the planetary pericentre and apocentre. Importantly, the net integrated torque curves reaffirm and validate the findings of \FRII whereupon the net torque undergoes a reversal when the characteristic ratio $e/h_{\rm p}$ exceeds unity and the planetary motion becomes supersonic. The quantitive correspondence between the simulations and linear theory is very good in the low eccentricity and transonic regime, with only small discrepancies arising for the highest values of $e$ considered. This agreement is also borne out in the orbital evolution rates as a function of $e$ where we recover the predicted $\tau_e^{-1} \propto e^{-3}$ dependency in the high $e$ limit. Meanwhile the $\tau_a^{-1}$ curve presents a more complex, non-monotonic structure, with outwards migration even possible in discs with shallow density profiles. Our consistent results caution against the use of previous fitting functions which present more notable discrepancies and are often limited to restricted regions of disc parameter space or $e$ intervals. The detailed comparison performed in this paper also acts as a gold standard for benchmarking new numerical setups for planet-disc interactions. Therefore we provide a repository of our linear theory data across a range of disc profiles and eccentricities, as described in appendix \ref{app:code_test}. 

Our simulations also allow us to probe fundamentally nonlinear effects which are not captured by the linear theory. In particular, the angular momentum flux profiles can differ as wave steepening and shocks deposit angular momentum into the disc. Whilst the peak deposition location is consistent with previous studies for near circular runs, for transonic and supersonic planets this shifts towards the apocentre and pericentre radii. The nonlinearities also allow us to comment on the long timescale development of coorbital horseshoe drags. At early times, our low mass ratio runs transiently develop the linear response and unsaturated corotation torque, although over long libration timescales we observe evidence for the onset of saturation. This temporal evolution of the coorbital angular momentum fluxes and torques becomes less pronounced as the eccentricity increases, supporting previous conclusions that the horseshoe drag is suppressed at high $e$.

Finally, we should acknowledge that despite the progress made within this study, it still leaves ample scope for further investigation. Here we have focused on 2D and locally isothermal discs. 3D effects and alternative thermodynamics are known to influence the planet-disc interaction problem. This promotes extensions to our global linear modal method and accompanying simulations, which will be explored in the future. 
% =========================== %
\section*{Acknowledgements}
% =========================== %

CWF and AJD would like to thank the referee for a detailed report which helped improve the clarity of this work. The authors would also like to acknowledge insightful conversations with Roman Rafikov, Yoram Lithwick, and Eugene Chiang during the preparation of this manuscript. We also thank Pablo Ben\'itez-Llambay for suggesting the inclusion of appendix \ref{app:code_test} as a benchmark for future numerical investigations. This research was supported by the W. M. Keck Foundation Fund and the IAS Fund for Memberships in the Natural Sciences (CWF), and by NASA through the NASA Hubble Fellowship grant
\#HST-HF2-51553.001 awarded by the Space Telescope Science Institute, which is operated by the Association of Universities for Research in Astronomy, Inc., for NASA, under contract NAS5-26555 (AJD). 

%%%%%%%%%%%%%%%%%%%%%%%%%%%%%%%%%%%%%%%%%%%%%%%%%%
\section*{Data Availability}

The data underlying this article are available in the online supplementary material.

%%%%%%%%%%%%%%%%%%%% REFERENCES %%%%%%%%%%%%%%%%%%

% The best way to enter references is to use BibTeX:

\bibliographystyle{mnras}
\bibliography{references.bib} % if your bibtex file is called example.bib

\begin{thebibliography}{}
\makeatletter
\relax
\def\mn@urlcharsother{\let\do\@makeother \do\$\do\&\do\#\do\^\do\_\do\%\do\~}
\def\mn@doi{\begingroup\mn@urlcharsother \@ifnextchar [ {\mn@doi@} {\mn@doi@[]}}
\def\mn@doi@[#1]#2{\def\@tempa{#1}\ifx\@tempa\@empty \href {http://dx.doi.org/#2} {doi:#2}\else \href {http://dx.doi.org/#2} {#1}\fi \endgroup}
\def\mn@eprint#1#2{\mn@eprint@#1:#2::\@nil}
\def\mn@eprint@arXiv#1{\href {http://arxiv.org/abs/#1} {{\tt arXiv:#1}}}
\def\mn@eprint@dblp#1{\href {http://dblp.uni-trier.de/rec/bibtex/#1.xml} {dblp:#1}}
\def\mn@eprint@#1:#2:#3:#4\@nil{\def\@tempa {#1}\def\@tempb {#2}\def\@tempc {#3}\ifx \@tempc \@empty \let \@tempc \@tempb \let \@tempb \@tempa \fi \ifx \@tempb \@empty \def\@tempb {arXiv}\fi \@ifundefined {mn@eprint@\@tempb}{\@tempb:\@tempc}{\expandafter \expandafter \csname mn@eprint@\@tempb\endcsname \expandafter{\@tempc}}}

\bibitem[\protect\citeauthoryear{Artymowicz}{Artymowicz}{1993}]{Artymowicz_1993}
Artymowicz P.,  1993, \mn@doi [\apj] {10.1086/173469}, 419, 155

\bibitem[\protect\citeauthoryear{{Artymowicz}, {Lin}  \& {Wampler}}{{Artymowicz} et~al.}{1993}]{1993ApJ...409..592A}
{Artymowicz} P.,  {Lin} D.~N.~C.,   {Wampler} E.~J.,  1993, \mn@doi [\apj] {10.1086/172690}, \href {https://ui.adsabs.harvard.edu/abs/1993ApJ...409..592A} {409, 592}

\bibitem[\protect\citeauthoryear{Arzamasskiy, Zhu  \& Stone}{Arzamasskiy et~al.}{2018}]{ArzamasskiyEtAl_2018}
Arzamasskiy L.,  Zhu Z.,   Stone J.~M.,  2018, \mn@doi [\mnras] {10.1093/mnras/sty001}, 475, 3201

\bibitem[\protect\citeauthoryear{Bellovary, Low, McKernan  \& Ford}{Bellovary et~al.}{2016}]{BellovaryEtAl_2016}
Bellovary J.~M.,  Low M.-M.~M.,  McKernan B.,   Ford K. E.~S.,  2016, \mn@doi [\apjl] {10.3847/2041-8205/819/2/L17}, 819, L17

\bibitem[\protect\citeauthoryear{Bitsch \& Kley}{Bitsch \& Kley}{2010}]{BitschKley_2010}
Bitsch B.,  Kley W.,  2010, \mn@doi [\aap] {10.1051/0004-6361/201014414}, 523, A30

\bibitem[\protect\citeauthoryear{Brown \& Ogilvie}{Brown \& Ogilvie}{2024}]{BrownOgilvie_2024a}
Brown J.~J.,  Ogilvie G.~I.,  2024, \mn@doi [\mnras] {10.1093/mnras/stae2060}, 534, 39

\bibitem[\protect\citeauthoryear{Cimerman \& Rafikov}{Cimerman \& Rafikov}{2021}]{CimermanRafikov_2021}
Cimerman N.~P.,  Rafikov R.~R.,  2021, \mn@doi [\mnras] {10.1093/mnras/stab2652}, 508, 2329

\bibitem[\protect\citeauthoryear{Cimerman, Rafikov  \& Miranda}{Cimerman et~al.}{2024}]{CimermanEtAl_2024}
Cimerman N.~P.,  Rafikov R.~R.,   Miranda R.,  2024, \mn@doi [\mnras] {10.1093/mnras/stae467}, 529, 425

\bibitem[\protect\citeauthoryear{Cordwell \& Rafikov}{Cordwell \& Rafikov}{2024}]{CordwellRafikov_2024}
Cordwell A.~J.,  Rafikov R.~R.,  2024, \mn@doi [\mnras] {10.1093/mnras/stae2089}, 534, 1394

\bibitem[\protect\citeauthoryear{Cresswell \& Nelson}{Cresswell \& Nelson}{2006}]{CresswellNelson_2006}
Cresswell P.,  Nelson R.~P.,  2006, \mn@doi [\aap] {10.1051/0004-6361:20054551}, 450, 833

\bibitem[\protect\citeauthoryear{Cresswell \& Nelson}{Cresswell \& Nelson}{2008}]{CresswellNelson_2008}
Cresswell P.,  Nelson R.~P.,  2008, \mn@doi [\aap] {10.1051/0004-6361:20079178}, 482, 677

\bibitem[\protect\citeauthoryear{Cresswell, Dirksen, Kley  \& Nelson}{Cresswell et~al.}{2007}]{CresswellEtAl_2007}
Cresswell P.,  Dirksen G.,  Kley W.,   Nelson R.~P.,  2007, \mn@doi [\aap] {10.1051/0004-6361:20077666}, 473, 329

\bibitem[\protect\citeauthoryear{D'Angelo \& Lubow}{D'Angelo \& Lubow}{2010}]{DAngeloLubow_2010}
D'Angelo G.,  Lubow S.~H.,  2010, \mn@doi [\apj] {10.1088/0004-637X/724/1/730}, 724, 730

\bibitem[\protect\citeauthoryear{Debras, Baruteau  \& Donati}{Debras et~al.}{2021}]{DebrasEtAl_2021}
Debras F.,  Baruteau C.,   Donati J.-F.,  2021, \mn@doi [\mnras] {10.1093/mnras/staa3397}, 500, 1621

\bibitem[\protect\citeauthoryear{Dempsey, Lee  \& Lithwick}{Dempsey et~al.}{2020}]{DempseyEtAl_2020}
Dempsey A.~M.,  Lee W.-K.,   Lithwick Y.,  2020, \mn@doi [\apj] {10.3847/1538-4357/ab723c}, 891, 108

\bibitem[\protect\citeauthoryear{{Derdzinski} \& {Mayer}}{{Derdzinski} \& {Mayer}}{2023}]{2023MNRAS.521.4522D}
{Derdzinski} A.,  {Mayer} L.,  2023, \mn@doi [\mnras] {10.1093/mnras/stad749}, \href {https://ui.adsabs.harvard.edu/abs/2023MNRAS.521.4522D} {521, 4522}

\bibitem[\protect\citeauthoryear{{Dittmann} \& {Miller}}{{Dittmann} \& {Miller}}{2020}]{2020MNRAS.493.3732D}
{Dittmann} A.~J.,  {Miller} M.~C.,  2020, \mn@doi [\mnras] {10.1093/mnras/staa463}, \href {https://ui.adsabs.harvard.edu/abs/2020MNRAS.493.3732D} {493, 3732}

\bibitem[\protect\citeauthoryear{{Duffell}}{{Duffell}}{2016}]{2016ApJS..226....2D}
{Duffell} P.~C.,  2016, \mn@doi [\apjs] {10.3847/0067-0049/226/1/2}, \href {https://ui.adsabs.harvard.edu/abs/2016ApJS..226....2D} {226, 2}

\bibitem[\protect\citeauthoryear{Duffell \& Chiang}{Duffell \& Chiang}{2015}]{DuffellChiang_2015}
Duffell P.~C.,  Chiang E.,  2015, \mn@doi [\apj] {10.1088/0004-637X/812/2/94}, 812, 94

\bibitem[\protect\citeauthoryear{Duffell \& MacFadyen}{Duffell \& MacFadyen}{2012}]{DuffellMacFadyen_2012}
Duffell P.~C.,  MacFadyen A.~I.,  2012, \mn@doi [\apj] {10.1088/0004-637X/755/1/7}, 755, 7

\bibitem[\protect\citeauthoryear{Eklund \& Masset}{Eklund \& Masset}{2017}]{EklundMasset_2017}
Eklund H.,  Masset F.~S.,  2017, \mn@doi [\mnras] {10.1093/mnras/stx856}, 469, 206

\bibitem[\protect\citeauthoryear{Eylen et~al.,}{Eylen et~al.}{2019}]{EylenEtAl_2019}
Eylen V.~V.,  et~al., 2019, \mn@doi [\aj] {10.3847/1538-3881/aaf22f}, 157, 61

\bibitem[\protect\citeauthoryear{Fairbairn \& Rafikov}{Fairbairn \& Rafikov}{2022}]{FairbairnRafikov_2022}
Fairbairn C.~W.,  Rafikov R.~R.,  2022, \mn@doi [\mnras] {10.1093/mnras/stac2802}, 517, 2121

\bibitem[\protect\citeauthoryear{Fairbairn \& Rafikov}{Fairbairn \& Rafikov}{2025}]{FairbairnRafikov_2025}
Fairbairn C.~W.,  Rafikov R.~R.,  2025, \mn@doi [\mnras] {10.1093/mnras/staf117}, 537, 1779

\bibitem[\protect\citeauthoryear{Fendyke \& Nelson}{Fendyke \& Nelson}{2014}]{FendykeNelson_2014}
Fendyke S.~M.,  Nelson R.~P.,  2014, \mn@doi [\mnras] {10.1093/mnras/stt1867}, 437, 96

\bibitem[\protect\citeauthoryear{Ford \& Rasio}{Ford \& Rasio}{2008}]{FordRasio_2008}
Ford E.~B.,  Rasio F.~A.,  2008, \mn@doi [\apj] {10.1086/590926}, 686, 621

\bibitem[\protect\citeauthoryear{{Frank}, {King}  \& {Raine}}{{Frank} et~al.}{2002}]{2002apa..book.....F}
{Frank} J.,  {King} A.,   {Raine} D.~J.,  2002, {Accretion Power in Astrophysics: Third Edition}

\bibitem[\protect\citeauthoryear{{Garg}, {Derdzinski}, {Zwick}, {Capelo}  \& {Mayer}}{{Garg} et~al.}{2022}]{2022MNRAS.517.1339G}
{Garg} M.,  {Derdzinski} A.,  {Zwick} L.,  {Capelo} P.~R.,   {Mayer} L.,  2022, \mn@doi [\mnras] {10.1093/mnras/stac2711}, \href {https://ui.adsabs.harvard.edu/abs/2022MNRAS.517.1339G} {517, 1339}

\bibitem[\protect\citeauthoryear{{Goldreich} \& {Sari}}{{Goldreich} \& {Sari}}{2003}]{Goldreich2003}
{Goldreich} P.,  {Sari} R.,  2003, \mn@doi [\apj] {10.1086/346202}, \href {https://ui.adsabs.harvard.edu/abs/2003ApJ...585.1024G} {585, 1024}

\bibitem[\protect\citeauthoryear{Goldreich \& Tremaine}{Goldreich \& Tremaine}{1979}]{GoldreichTremaine_1979}
Goldreich P.,  Tremaine S.,  1979, \mn@doi [\apj] {10.1086/157448}, 233, 857

\bibitem[\protect\citeauthoryear{Goldreich \& Tremaine}{Goldreich \& Tremaine}{1980}]{GoldreichTremaine_1980}
Goldreich P.,  Tremaine S.,  1980, \mn@doi [\apj] {10.1086/158356}, \href {https://ui.adsabs.harvard.edu/abs/1980ApJ...241..425G} {241, 425}

\bibitem[\protect\citeauthoryear{Goodman \& Rafikov}{Goodman \& Rafikov}{2001}]{GoodmanRafikov_2001}
Goodman J.,  Rafikov R.~R.,  2001, \mn@doi [\apj] {10.1086/320572}, 552, 793

\bibitem[\protect\citeauthoryear{{Gottlieb} \& {Shu}}{{Gottlieb} \& {Shu}}{1998}]{1998MaCom..67...73G}
{Gottlieb} S.,  {Shu} C.~W.,  1998, Mathematics of Computation, \href {https://ui.adsabs.harvard.edu/abs/1998MaCom..67...73G} {67, 73}

\bibitem[\protect\citeauthoryear{Ida, Muto, Matsumura  \& Brasser}{Ida et~al.}{2020}]{IdaEtAl_2020}
Ida S.,  Muto T.,  Matsumura S.,   Brasser R.,  2020, \mn@doi [\mnras] {10.1093/mnras/staa1073}, 494, 5666

\bibitem[\protect\citeauthoryear{Jim{\'e}nez \& Masset}{Jim{\'e}nez \& Masset}{2017}]{JimenezMasset_2017}
Jim{\'e}nez M.~A.,  Masset F.~S.,  2017, \mn@doi [\mnras] {10.1093/mnras/stx1946}, 471, 4917

\bibitem[\protect\citeauthoryear{Kane, Ciardi, Gelino  \& {von Braun}}{Kane et~al.}{2012}]{KaneEtAl_2012}
Kane S.~R.,  Ciardi D.~R.,  Gelino D.~M.,   {von Braun} K.,  2012, \mn@doi [\mnras] {10.1111/j.1365-2966.2012.21627.x}, 425, 757

\bibitem[\protect\citeauthoryear{Kley, Müller, Kolb, Benítez-Llambay  \& Masset}{Kley et~al.}{2012}]{KleyEtAl_2012}
Kley W.,  Müller T. W.~A.,  Kolb S.~M.,  Benítez-Llambay P.,   Masset F.,  2012, \mn@doi [\aap] {10.1051/0004-6361/201219719}, 546, A99

\bibitem[\protect\citeauthoryear{Korycansky \& Pollack}{Korycansky \& Pollack}{1993}]{KorycanskyPollack_1993}
Korycansky D.,  Pollack J.,  1993, \mn@doi [Icarus] {10.1006/icar.1993.1039}, 102, 150

\bibitem[\protect\citeauthoryear{{Kurganov} \& {Tadmor}}{{Kurganov} \& {Tadmor}}{2000}]{2000JCoPh.160..241K}
{Kurganov} A.,  {Tadmor} E.,  2000, \mn@doi [Journal of Computational Physics] {10.1006/jcph.2000.6459}, \href {https://ui.adsabs.harvard.edu/abs/2000JCoPh.160..241K} {160, 241}

\bibitem[\protect\citeauthoryear{Lega, Morbidelli  \& Nesvorný}{Lega et~al.}{2013}]{LegaEtAl_2013}
Lega E.,  Morbidelli A.,   Nesvorný D.,  2013, \mn@doi [\mnras] {10.1093/mnras/stt431}, 431, 3494

\bibitem[\protect\citeauthoryear{{Lynden-Bell} \& Pringle}{{Lynden-Bell} \& Pringle}{1974}]{Lynden-BellPringle_1974}
{Lynden-Bell} D.,  Pringle J.~E.,  1974, \mn@doi [\mnras] {10.1093/mnras/168.3.603}, 168, 603

\bibitem[\protect\citeauthoryear{Masset}{Masset}{2017}]{Masset_2017}
Masset F.~S.,  2017, \mn@doi [\mnras] {10.1093/mnras/stx2271}, 472, 4204

\bibitem[\protect\citeauthoryear{{McKernan}, {Ford}  \& {O'Shaughnessy}}{{McKernan} et~al.}{2020}]{2020MNRAS.498.4088M}
{McKernan} B.,  {Ford} K.~E.~S.,   {O'Shaughnessy} R.,  2020, \mn@doi [\mnras] {10.1093/mnras/staa2681}, \href {https://ui.adsabs.harvard.edu/abs/2020MNRAS.498.4088M} {498, 4088}

\bibitem[\protect\citeauthoryear{Miranda \& Rafikov}{Miranda \& Rafikov}{2019a}]{MirandaRafikov_2019a}
Miranda R.,  Rafikov R.~R.,  2019a, \mn@doi [\apj] {10.3847/1538-4357/ab0f9e}, 875, 37

\bibitem[\protect\citeauthoryear{{Miranda} \& {Rafikov}}{{Miranda} \& {Rafikov}}{2019b}]{MirandaRafikov_2019b}
{Miranda} R.,  {Rafikov} R.~R.,  2019b, \mn@doi [\apjl] {10.3847/2041-8213/ab22a7}, \href {https://ui.adsabs.harvard.edu/abs/2019ApJ...878L...9M} {878, L9}

\bibitem[\protect\citeauthoryear{Miranda \& Rafikov}{Miranda \& Rafikov}{2020}]{MirandaRafikov_2020}
Miranda R.,  Rafikov R.~R.,  2020, \mn@doi [\apj] {10.3847/1538-4357/ab791a}, 892, 65

\bibitem[\protect\citeauthoryear{Monnier et~al.,}{Monnier et~al.}{2019}]{MonnierEtAl_2019}
Monnier J.~D.,  et~al., 2019, \mn@doi [\apj] {10.3847/1538-4357/aafe87}, 872, 122

\bibitem[\protect\citeauthoryear{Murray \& Dermott}{Murray \& Dermott}{1999}]{MurrayDermott_1999}
Murray C.~D.,  Dermott S.~F.,  1999, Solar system dynamics.
\url {https://ui.adsabs.harvard.edu/abs/1999ssd..book.....M}

\bibitem[\protect\citeauthoryear{Muto, Takeuchi  \& Ida}{Muto et~al.}{2011}]{MutoEtAl_2011}
Muto T.,  Takeuchi T.,   Ida S.,  2011, \mn@doi [\apj] {10.1088/0004-637X/737/1/37}, 737, 37

\bibitem[\protect\citeauthoryear{O'Neill, D'Orazio, Samsing  \& Pessah}{O'Neill et~al.}{2024}]{ONeillEtAl_2024}
O'Neill D.,  D'Orazio D.~J.,  Samsing J.,   Pessah M.~E.,  2024, \mn@doi [\apj] {10.3847/1538-4357/ad7250}, 974, 216

\bibitem[\protect\citeauthoryear{Ogilvie \& Lubow}{Ogilvie \& Lubow}{2003}]{OgilvieLubow_2003}
Ogilvie G.~I.,  Lubow S.~H.,  2003, \mn@doi [\apj] {10.1086/368178}, 587, 398

\bibitem[\protect\citeauthoryear{Ogilvie, Latter  \& Lesur}{Ogilvie et~al.}{2025}]{OgilvieEtAl_2025}
Ogilvie G.~I.,  Latter H.~N.,   Lesur G.,  2025, \mn@doi [\mnras] {10.1093/mnras/staf154}, 537, 3349

\bibitem[\protect\citeauthoryear{Ostriker}{Ostriker}{1999}]{Ostriker_1999}
Ostriker E.~C.,  1999, \mn@doi [\apj] {10.1086/306858}, 513, 252

\bibitem[\protect\citeauthoryear{Paardekooper \& Papaloizou}{Paardekooper \& Papaloizou}{2009}]{PaardekooperPapaloizou_2009a}
Paardekooper S.-J.,  Papaloizou J. C.~B.,  2009, \mn@doi [\mnras] {10.1111/j.1365-2966.2009.14511.x}, 394, 2283

\bibitem[\protect\citeauthoryear{Paardekooper, Baruteau, Crida  \& Kley}{Paardekooper et~al.}{2010}]{PaardekooperEtAl_2010}
Paardekooper S.~J.,  Baruteau C.,  Crida A.,   Kley W.,  2010, \mn@doi [\mnras] {10.1111/j.1365-2966.2009.15782.x}, 401, 1950

\bibitem[\protect\citeauthoryear{{Paardekooper}, {Dong}, {Duffell}, {Fung}, {Masset}, {Ogilvie}  \& {Tanaka}}{{Paardekooper} et~al.}{2023}]{Paardekooper2023}
{Paardekooper} S.,  {Dong} R.,  {Duffell} P.,  {Fung} J.,  {Masset} F.~S.,  {Ogilvie} G.,   {Tanaka} H.,  2023, in {Inutsuka} S.,  {Aikawa} Y.,  {Muto} T.,  {Tomida} K.,   {Tamura} M.,  eds,  Astronomical Society of the Pacific Conference Series Vol. 534, Protostars and Planets VII. p.~685 (\mn@eprint {arXiv} {2203.09595}), \mn@doi{10.48550/arXiv.2203.09595}

\bibitem[\protect\citeauthoryear{Papaloizou \& Larwood}{Papaloizou \& Larwood}{2000}]{PapaloizouLarwood_2000}
Papaloizou J. C.~B.,  Larwood J.~D.,  2000, \mn@doi [\mnras] {10.1046/j.1365-8711.2000.03466.x}, 315, 823

\bibitem[\protect\citeauthoryear{{Petrovich} \& {Rafikov}}{{Petrovich} \& {Rafikov}}{2012}]{Petrovich2012}
{Petrovich} C.,  {Rafikov} R.~R.,  2012, \mn@doi [\apj] {10.1088/0004-637X/758/1/33}, \href {https://ui.adsabs.harvard.edu/abs/2012ApJ...758...33P} {758, 33}

\bibitem[\protect\citeauthoryear{Pichierri, Bitsch  \& Lega}{Pichierri et~al.}{2023}]{PichierriEtAl_2023}
Pichierri G.,  Bitsch B.,   Lega E.,  2023, \mn@doi [\aap] {10.1051/0004-6361/202245196}, 670, A148

\bibitem[\protect\citeauthoryear{Pichierri, Bitsch  \& Lega}{Pichierri et~al.}{2024}]{PichierriEtAl_2024}
Pichierri G.,  Bitsch B.,   Lega E.,  2024, \mn@doi [\apj] {10.3847/1538-4357/ad3dff}, 967, 111

\bibitem[\protect\citeauthoryear{{Rafikov}}{{Rafikov}}{2002a}]{Rafikov2002a}
{Rafikov} R.~R.,  2002a, \mn@doi [\apj] {10.1086/339399}, \href {https://ui.adsabs.harvard.edu/abs/2002ApJ...569..997R} {569, 997}

\bibitem[\protect\citeauthoryear{Rafikov}{Rafikov}{2002b}]{Rafikov_2002}
Rafikov R.~R.,  2002b, \mn@doi [\apj] {10.1086/340228}, 572, 566

\bibitem[\protect\citeauthoryear{Rafikov \& Petrovich}{Rafikov \& Petrovich}{2012}]{RafikovPetrovich_2012}
Rafikov R.~R.,  Petrovich C.,  2012, \mn@doi [\apj] {10.1088/0004-637X/747/1/24}, 747, 24

\bibitem[\protect\citeauthoryear{{Ragusa}, {Rosotti}, {Teyssandier}, {Booth}, {Clarke}  \& {Lodato}}{{Ragusa} et~al.}{2018}]{Ragusa2018}
{Ragusa} E.,  {Rosotti} G.,  {Teyssandier} J.,  {Booth} R.,  {Clarke} C.~J.,   {Lodato} G.,  2018, \mn@doi [\mnras] {10.1093/mnras/stx3094}, \href {https://ui.adsabs.harvard.edu/abs/2018MNRAS.474.4460R} {474, 4460}

\bibitem[\protect\citeauthoryear{R{\"o}pke \& De~Marco}{R{\"o}pke \& De~Marco}{2023}]{RopkeDeMarco_2023}
R{\"o}pke F.~K.,  De~Marco O.,  2023, \mn@doi [Living Reviews in Computational Astrophysics] {10.1007/s41115-023-00017-x}, 9, 2

\bibitem[\protect\citeauthoryear{{Stone}, {Metzger}  \& {Haiman}}{{Stone} et~al.}{2017}]{2017MNRAS.464..946S}
{Stone} N.~C.,  {Metzger} B.~D.,   {Haiman} Z.,  2017, \mn@doi [\mnras] {10.1093/mnras/stw2260}, \href {https://ui.adsabs.harvard.edu/abs/2017MNRAS.464..946S} {464, 946}

\bibitem[\protect\citeauthoryear{{Tagawa}, {Haiman}  \& {Kocsis}}{{Tagawa} et~al.}{2020}]{2020ApJ...898...25T}
{Tagawa} H.,  {Haiman} Z.,   {Kocsis} B.,  2020, \mn@doi [\apj] {10.3847/1538-4357/ab9b8c}, \href {https://ui.adsabs.harvard.edu/abs/2020ApJ...898...25T} {898, 25}

\bibitem[\protect\citeauthoryear{{Takeuchi}, {Miyama}  \& {Lin}}{{Takeuchi} et~al.}{1996}]{Takeuchi1996}
{Takeuchi} T.,  {Miyama} S.~M.,   {Lin} D.~N.~C.,  1996, \mn@doi [\apj] {10.1086/177013}, \href {https://ui.adsabs.harvard.edu/abs/1996ApJ...460..832T} {460, 832}

\bibitem[\protect\citeauthoryear{Tanaka \& Okada}{Tanaka \& Okada}{2024}]{TanakaOkada_2024}
Tanaka H.,  Okada K.,  2024, \mn@doi [\apj] {10.3847/1538-4357/ad410d}, 968, 28

\bibitem[\protect\citeauthoryear{Tanaka \& Ward}{Tanaka \& Ward}{2004}]{TanakaWard_2004}
Tanaka H.,  Ward W.~R.,  2004, \mn@doi [\apj] {10.1086/380992}, 602, 388

\bibitem[\protect\citeauthoryear{Tanaka, Takeuchi  \& Ward}{Tanaka et~al.}{2002}]{TanakaEtAl_2002}
Tanaka H.,  Takeuchi T.,   Ward W.~R.,  2002, \mn@doi [\apj] {10.1086/324713}, 565, 1257

\bibitem[\protect\citeauthoryear{Teyssandier \& Ogilvie}{Teyssandier \& Ogilvie}{2016}]{TeyssandierOgilvie_2016}
Teyssandier J.,  Ogilvie G.~I.,  2016, \mn@doi [\mnras] {10.1093/mnras/stw521}, 458, 3221

\bibitem[\protect\citeauthoryear{{Toro}, {Spruce}  \& {Speares}}{{Toro} et~al.}{1994}]{1994ShWav...4...25T}
{Toro} E.~F.,  {Spruce} M.,   {Speares} W.,  1994, \mn@doi [Shock Waves] {10.1007/BF01414629}, \href {https://ui.adsabs.harvard.edu/abs/1994ShWav...4...25T} {4, 25}

\bibitem[\protect\citeauthoryear{Uyama et~al.,}{Uyama et~al.}{2020}]{UyamaEtAl_2020}
Uyama T.,  et~al., 2020, \mn@doi [\apj] {10.3847/1538-4357/aba8f6}, 900, 135

\bibitem[\protect\citeauthoryear{Ward}{Ward}{1986}]{Ward_1986}
Ward R.,  1986, Icarus

\bibitem[\protect\citeauthoryear{Ward}{Ward}{1991}]{Ward_1991}
Ward W.~R.,  1991, Conference Name: Lunar and Planetary Science Conference, 22, 1463

\bibitem[\protect\citeauthoryear{Xie et~al.,}{Xie et~al.}{2016}]{XieEtAl_2016}
Xie J.-W.,  et~al., 2016, \mn@doi [Proceedings of the National Academy of Sciences] {10.1073/pnas.1604692113}, 113, 11431

\bibitem[\protect\citeauthoryear{Xie, Ren, Dong, Pueyo, Ruffio, Fang, Mawet  \& Stolker}{Xie et~al.}{2021}]{XieEtAl_2021}
Xie C.,  Ren B.,  Dong R.,  Pueyo L.,  Ruffio J.-B.,  Fang T.,  Mawet D.,   Stolker T.,  2021, \mn@doi [\apj] {10.3847/2041-8213/abd241}, 906, L9

\bibitem[\protect\citeauthoryear{{Yunes}, {Kocsis}, {Loeb}  \& {Haiman}}{{Yunes} et~al.}{2011}]{2011PhRvL.107q1103Y}
{Yunes} N.,  {Kocsis} B.,  {Loeb} A.,   {Haiman} Z.,  2011, \mn@doi [\prl] {10.1103/PhysRevLett.107.171103}, \href {https://ui.adsabs.harvard.edu/abs/2011PhRvL.107q1103Y} {107, 171103}

\bibitem[\protect\citeauthoryear{Zhu \& Zhang}{Zhu \& Zhang}{2022}]{ZhuZhang_2022}
Zhu Z.,  Zhang R.~M.,  2022, \mn@doi [\mnras] {10.1093/mnras/stab3641}, 510, 3986

\bibitem[\protect\citeauthoryear{{van Leer}}{{van Leer}}{1979}]{1979JCoPh..32..101V}
{van Leer} B.,  1979, \mn@doi [Journal of Computational Physics] {10.1016/0021-9991(79)90145-1}, \href {https://ui.adsabs.harvard.edu/abs/1979JCoPh..32..101V} {32, 101}

\makeatother
\end{thebibliography}

%%%%%%%%%%%%%%%%%%%%%%%%%%%%%%%%%%%%%%%%%%%%%%%%%%

%%%%%%%%%%%%%%%%% APPENDICES %%%%%%%%%%%%%%%%%%%%%

\appendix

% ================================%
\section{Convergence tests}
\label{app:convergence}
% ================================%

In this section we will interrogate the convergence properties for both our linear framework and numerical simulations. In section \ref{subsec:linear_convergence} we will examine how our linear framework converges with the number of included modes. Meanwhile in section \ref{subsec:simulation_convergence} we will test the effect of resolution on our simulation. 

% ...........................................%
\subsection{Linear theory modal convergence}
\label{subsec:linear_convergence}
% ...........................................%

\begin{figure}
    \centering
    \includegraphics[width=\linewidth]{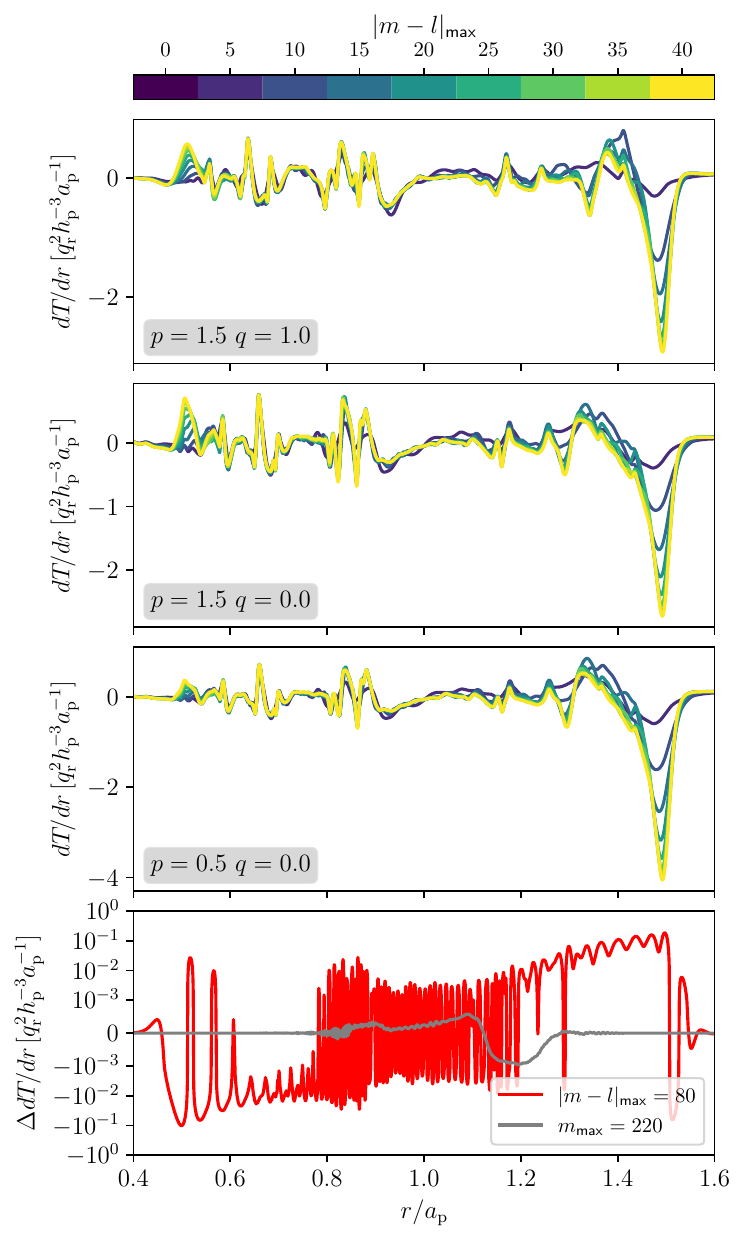}
    \caption{\textit{Upper three panels}: convergence study of $dT/dr$ for our three disc profiles and $e=0.5$. The solid lines denote the profiles holding $m_{\rm max} = 170$ fixed and varying $|m-l|_{\rm max}$ between 5 (purple) and 40 (yellow). \textit{Lower panel}: the torque density residuals between the fiducial modal range and the extended sums with $|m-l|_{\rm max} =80$ (red line) and $m_{\rm max} = 220$ (grey line), for $e = 0.5$ in the $p=0.5$ disc model.}
    \label{fig:dTdr_convergence}
\end{figure}

In this section, we will interrogate the modal convergence of our linear methodology. Indeed, since this paper considers highly eccentric orbits, beyond the scope which was confidently converged in the previous work of \FRII, we anticipate the need for a high number of modes. Throughout the main body of this paper, we have assumed a default modal range of $1 < m < 170$ and $|m-l| \leq 40$. In the upper three panels of Fig.~\ref{fig:dTdr_convergence}, we show the radial torque density for our three disc parameter combinations whilst adopting the particular value of $e = 0.5$. This reference eccentricity is chosen since more noticeable discrepancies begin to arise between the linear framework and simulations beyond this value (see Fig.~\ref{fig:torque_pq} for example). The convergence properties for this high value of $e$ will bound faster convergence behaviour for smaller values of $e$, whilst the larger values are expected to converge more slowly. In each panel, the coloured lines correspond to the profiles obtained with different maximal values of $|m-l|_{\rm max}$, whilst $m_{\rm max} = 170$ is held fixed at the standard value. The results span from $|m-l|_{\rm} = 5$  (dark purple line), up to the fiducial $|m-l|_{\rm max} = 40$  (yellow line), in increments of 5. Note that we restrict our attention to the radial range $0.4< r/a < 1.6$. Outside of this interval, the torque density is of lower amplitude and the torque wiggle features are predominantly formed by the action of low order modes in $m$ and $|m-l|$ (see discussion in section \ref{subsec:torque_profiles}).

One clearly observes a change of the torque profiles as more off-diagonal $|m-l|$ modes are included. The maximum peak and minimum trough, located at peri and apocentre respectively, are underestimated and less well defined when fewer modes are included. Intuitively, \cite{MurrayDermott_1999} show that the disturbing function Fourier components for eccentric perturbers are proportional to $e^{|m-l|}$. Therefore, for larger $e$, we expect the increasing importance of modal combinations with $l \neq m$. By the point that $|m-l|_{\rm max} \geq 35$ the profiles seem rather well converged, with the main features established and only very subtle changes as the included off-diagonal modes are expanded further. 

Indeed, to check this, we additionally compute the results for two extended modal intervals in our $p=0.5$ disc -- one with $|m-l|_{\rm max} = 80$ whilst keeping $m_{\rm max} = 170$, and the other with the fiducial $|m-l|_{\rm max} = 40$ but increased $m_{\rm max} = 220$. In the bottom panel of Fig.~\ref{fig:dTdr_convergence} we then plot the residual between these extended runs and the fiducial modal range,
\begin{equation}
    \Delta \frac{dT}{dr} = \left.\frac{dT}{dr}\right|_{\rm fiducial}-\left.\frac{dT}{dr}\right|_{\rm extended}.
\end{equation}
We see that the extended range in $m_{\rm max}$ (grey line) presents very small residuals which contribute only at the order of $\sim 10^{-3}$. Meanwhile, the wider $|m-l|_{\rm max}$ (red line) leads to more prominent corrections at the level of $\sim 10^{-1}$, but these also are very weak relative to the net signal which will barely change.

\begin{figure}
    \centering
    \includegraphics[width=\linewidth]{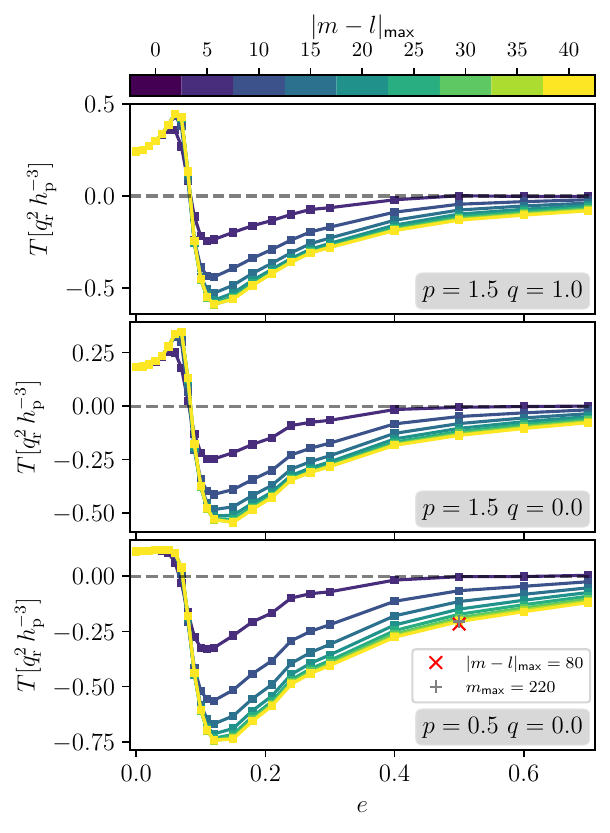}
    \caption{Convergence study for the net integrated torque across our three disc profiles and $e$ values. The square markers and connective lines show the $T$ profiles as functions of $e$, for fixed $m_{\rm max}=170$ and increasing values of $|m-l|_{\rm max}$ from 5 (purple) to 40 (yellow). In the $p=0.5$ panel, we additionally plot the $e = 0.5$ data point for the extended modal ranges with $(m_{\rm max}, |l-m|_{\rm max}) = (170,80)$ [red $\times$] and (220,40) [grey $+$].}
    \label{fig:net_T_converegence}
\end{figure}

In Fig.~\ref{fig:net_T_converegence} we now examine the convergence of the integrated torque for the full range of $e$. Akin to Fig.~\ref{fig:torque_pq}, we plot the net integrated torque curves as a function of $e$ for our three fiducial disc setups. Once again, the coloured lines denote the linear calculation results when including an increasing number of $|m-l|$ modes, from 5 (dark purple) up to 40 (yellow). All three panels tell a similar story. For sufficiently low values of $e$, well below the transonic transition and torque reversal, all of the curves overlap and a low number of off-diagonal modes satisfy convergence. Of course, for $e=0.0$ only the diagonal $m=l$ modes feature anyway. Pushing to higher values of $e$, we observe that the magnitude of the torque is severely underestimated unless a sufficient number of modes are included. The discrepancies are most pronounced just beyond the transonic turnover, where the net torque reaches an absolute minimum. It seems that for $|m-l|_{\rm max} = 25$ and above, the curves all overlap and are sufficiently converged. Note that this convergence in the integrated net torque is slightly less stringent than the detailed convergence in the full radial torque density profile. 

Once again, in order to further scrutinise the robustness of this fiducial modal range, we also plot the data points for the extended mode summation, as considered above for the disc with $p=0.5$ and $e = 0.5$. The $|m-l|_{\rm max} = 80$ and $m_{\rm max} = 220$ markers closely overly the fiducial data point, once again indicating that the inclusion of a reasonable number of additional modes incurs no significant changes in the linear theory torque diagnostics. More quantitatively, radially integrating over the residuals $\Delta dT/dr$ seen in the lower panel of Fig.~\ref{fig:dTdr_convergence}, gives a net discrepancy of $0.011$  and $-2.1\times10^{-5}$ for the extended $|m-l|_{\rm max}$ and $m_{\rm max}$ runs respectively. Both of these corrections are still too small to bring the simulations and linear theory into perfect alignment which differ by $\sim 0.06$ for the fiducial modal sum at this specific data point.

% ...........................................%
\subsection{Simulation resolution convergence}
\label{subsec:simulation_convergence}
% ...........................................%

In addition to testing the  modal convergence of the linear theory, it is also prudent to check that the torque diagnostics from our numerical simulations are well converged with increasing resolution. To this end, we have performed a high-resolution run targeting the parameters $p=0.5$, $q = 0.0$ and $e=0.5$, for which we see torque deviations beginning to emerge between our fiducial simulations and linear theory in Fig.~\ref{fig:torque_pq}. We adopt double the fiducial resolution described in section \ref{sec:numerical_setup}, such that $N_{r} = 9216$ and $N_{\phi}\approx13058$.

\begin{figure}
    \centering
    \includegraphics[width=\linewidth]{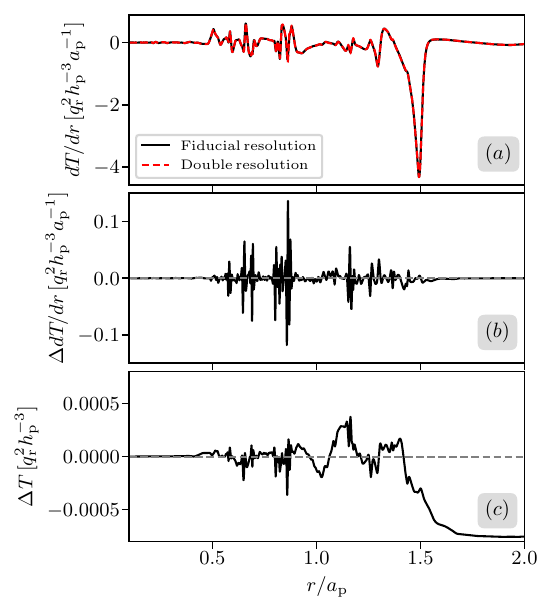}
    \caption{Resolution convergence test for $(p,q) = (0.5,0.0)$ and $e=0.5$ \texttt{Disco} simulations. (a) Compares the radial torque density for our fiducial (black solid) and double resolution (red dashed) runs. (b) Computes the residual torque density between these two runs. (c) Shows the cumulative torque error between the two runs integrated out from $r = 0.1 a_{\rm p}$.  }
    \label{fig:dTdr_convergence_sim}
\end{figure}

In Fig.~\ref{fig:dTdr_convergence_sim} we then compare the results for our fiducial and double resolution experiments. In panel (a) we plot the radial torque density for our standard (black solid) and double resolution (red dashed) runs, focusing on the radial interval between pericentre and apocentre where most of the complex structure is located. The curves effectively overlap, with only very small differences which are imperceptible on this scale. In panel (b) we isolate these differences by constructing the residual between the two torque curves
\begin{equation}
    \Delta \frac{dT}{dr} = \left.\frac{dT}{dr}\right|_{\rm fiducial}-\left.\frac{dT}{dr}\right|_{\rm high\,res}.
\end{equation}
This highlights the presence of small discrepancies, typically co-located with the spiky peaks and troughs of the $dT/dr$ profiles. The relevance of these small differences are further suppressed when one integrates over the residual, since they tend to cancel each other out. The cumulative torque error, integrating out from $r = 0.1 a_{\rm p}$, is then shown in panel (c). This error remains very small and by $r = 2.0 a_{\rm p}$ begins to level off around $\sim 7.5\times10^{-4}$. This is clearly far too small to bring the fiducial simulations and linear framework into alignment, since they exhibit a much more prominent discrepancy in $T$ at the level of $\sim 0.06$.

% ...........................................%
\subsection{Convergence remarks}
\label{subsec:convergence_remarks}
% ...........................................%

In summary, the tests performed in sections \ref{subsec:linear_convergence} and \ref{subsec:simulation_convergence} suggest that the torque diagnostics explored in our main results are suitably converged in terms of the linear modal range and numerical resolution.\footnote{Note however that we required a higher resolution to capture the converged nonlinear shocking behaviour discussed in section \ref{subsec:amf_deposit}.} Furthermore, the very close agreement between the linear framework and simulations below $e=0.5$ lends confidence to this conclusion.  

However, we do acknowledge the emerging differences between the linear theory and simulation results beyond $e = 0.5$. Since the inclusion of more modes and doubled resolution do not produce an appreciable change, this could point towards more fundamental difficulties in capturing the launching of waves in both the linear theory and/or simulations at high $e$. 

\begin{figure}
    \centering
    \includegraphics[width=0.96\linewidth]{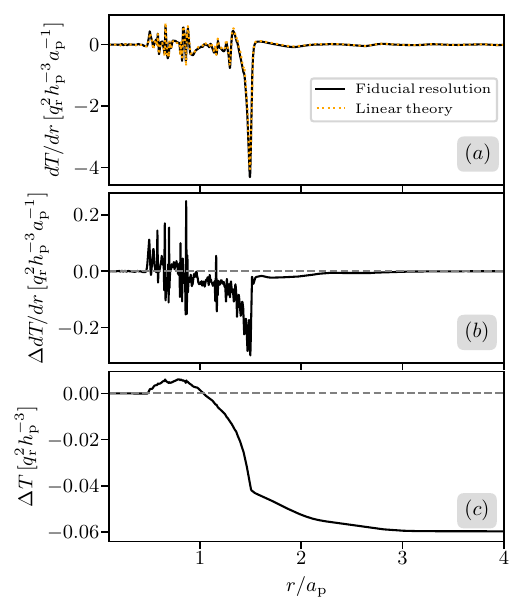}
    \caption{Same as for Fig.~\ref{fig:dTdr_convergence_sim} but now comparing the residual discrepancies between the fiducial linear and numerical approaches.}
    \label{fig:dTdr_err}
\end{figure}

In Fig.~\ref{fig:dTdr_err} we examine in more detail where these discrepancies are accrued. This plot is analogous to Fig.~\ref{fig:dTdr_convergence_sim} but now we are comparing the fiducial simulation and linear theory results. Panel (a) once again demonstrates the broadly good agreement for the main features in the torque density profile, as was previously plotted in the bottom panel of Fig.~\ref{fig:torque_p0.5q0.0}. Looking at the discrepancy in panel (b), $dT/dr|_{\rm sim}-dT/dr|_{\rm lin}$, we see that the errors are concentrated between the pericentre and apocentre radii. Indeed, integrating this residual profile yields the cumulative torque error shown in panel (c), which highlights that the dominant discrepancy manifests about the outer portion of the eccentric orbit, near apocentre.

Whilst one might suspect that the torque differences could be attributed to boundary conditions problems, as the highly eccentric planet moves ever closer to the edge, here we see that the errors instead build up between the radial excursions of the planet. The complex wake structure in this region might be susceptible to other (weakly) nonlinear effects which could modify the torques found by \texttt{Disco}. Alternatively, these sharp and intricate features might simply converge too slowly in our linear theory, requiring an unfeasible amount of modes. Finally, as we have seen in section \ref{subsec:amf_deposit}, the deposition of torque is non-zero between apocentre and pericentre for highly eccentric perturbers. This may modify the background surface density profile in this region and therefore alter the exchange of angular momentum between the planet and the disc. Exploring the origin of this discrepancy further will be pursued in future work.

% ================================%
\section{A New Benchmark for planet-disc numerics}
\label{app:code_test}
% ================================%

The focused study presented in this paper demonstrates the remarkable agreement that can be obtained between linear theory and numerical simulations, when appropriate care is taken. Having established this effective correspondence, we promote the use of our linear theory calculations as a controlled benchmark for testing future planet-disc interaction problems. Indeed, a detailed code versus theory comparison in the linear regime might also lend confidence to the numerical scheme as more general setups are considered (e.g. higher planet masses, different thermal prescriptions, viscous discs). To this end, we have provided a supplementary data repository which contains our torque density and AMF profiles across the three disc parameters combinations and a range of eccentricities, $e \in (0.01,0.06,0.12,0.3)$. This is accompanied by exemplary reading and plotting scripts which show how to access and visualise the data. Together, these offer a new standard benchmark for testing the performance of planet-disc problems. The repository can be accessed at \url{https://github.com/callumf8/linear_ecc_planet}.

%%%%%%%%%%%%%%%%%%%%%%%%%%%%%%%%%%%%%%%%%%%%%%%%%%

% Don't change these lines
\bsp	% typesetting comment
\label{lastpage}
\end{document}